\documentclass[draft]{agujournal2019}
\usepackage{url} 
\usepackage{lineno}
\usepackage[inline]{trackchanges} 
\usepackage{soul}
\usepackage{natbib}
\usepackage{hyperref}
\usepackage{multirow}
\usepackage{lscape}
\usepackage{amsmath}
\usepackage{array}
\linenumbers
\draftfalse

\journalname{Space Weather}

\begin{document}

\nolinenumbers
%
%


\title{
Conductance Model for Extreme Events : Impact of Auroral Conductance on Space Weather Forecasts}

%
%




\authors{Agnit Mukhopadhyay\affil{1}, Daniel T. Welling\affil{2}, Michael W. Liemohn\affil{1}, Aaron J. Ridley\affil{1}, Shibaji Chakraborty\affil{3}, and Brian J. Anderson\affil{4}}


\affiliation{1}{Climate and Space Sciences and Engineering Department, University of Michigan, Ann Arbor, MI, USA}
\affiliation{2}{Department of Physics, University of Texas at Arlington, Arlington, TX, USA}
\affiliation{3}{Department of Electrical and Computer Engineering, Virginia Polytechnic Institute and State University, Blacksburg, VA, USA}
\affiliation{4}{Applied Physics Laboratory, Johns Hopkins University, Baltimore, MD, USA}




\correspondingauthor{Agnit Mukhopadhyay}{agnitm@umich.edu}




\begin{keypoints}
\item An updated auroral conductance module 
is built for global models using nonlinear regression \& empirical adjustments spanning extreme events.
\item Expanded dataset raises the ceiling of conductance values, impacting the polar cap potential, $dB/dt$ \& $\Delta B$ predictions during extreme events.
\item Application of expanded model with empirical oval 
adjustments refines the conductance pattern, and drastically improves $dB/dt$ predictions. 

\end{keypoints}

%
%

%
%


\begin{abstract}
{{Ionospheric conductance is a crucial factor in regulating the closure of magnetospheric field-aligned currents through the ionosphere as Hall and Pedersen currents.}}
Despite its importance in predictive investigations of the magnetosphere - ionosphere coupling, the estimation of ionospheric conductance in the auroral region is precarious in most global first-principles based models.
{{This impreciseness in estimating the auroral conductance impedes both our understanding and predictive capabilities of the magnetosphere-ionosphere system during extreme space weather events. }}
In this article, we address this concern, with the development of an advanced Conductance Model for Extreme Events (CMEE) that estimates the auroral conductance from field aligned current values. 
CMEE has been developed using nonlinear regression over a year's worth of one-minute resolution output from assimilative maps, specifically including times of extreme driving of the solar wind-magnetosphere-ionosphere system. 
The model also includes provisions to enhance the conductance in the aurora using additional adjustments to refine the auroral oval.
CMEE has been incorporated within the Ridley Ionosphere Model (RIM) of the Space Weather Modeling Framework (SWMF) for usage in space weather simulations. This paper compares performance of CMEE against the existing conductance model in RIM, through a validation process for six space weather events. {{The performance analysis indicates overall improvement in the ionospheric feedback to ground-based space weather forecasts.}} Specifically, the model is able to improve the prediction of ionospheric currents which impact the simulated $dB/dt$ and $\Delta B$, resulting in substantial improvements in $dB/dt$ predictive skill. 


\end{abstract}

\section*{Plain Language Summary}
Electric currents generated in the Earth's space environment due to its magnetic interaction with the Sun leads to charged particle deposition and closure of these currents in the terrestrial upper atmosphere, especially in the high latitude auroral region. The enhancement in the electrical charge carrying capacity as a result of this process in the Earth's upper atmosphere, also known as the ionosphere, is challenging to estimate in most numerical simulations attempting to study the interactive 
dynamic and chemical processes in the near-Earth region. The inability to accurately estimate this quantity leads to underprediction of severe space weather events that can have adverse impacts on man-made technology like electrical power grids, railway and oil pipelines. In this study, we present a novel modeling approach to address this problem, and provide global simulations with a more accurate estimate on the electrical conductivity of the ionosphere. Through this investigation, we show that the accurate measurement of the charge carriers in the ionosphere using the new model causes substantial improvements in the prediction of space weather on the ground, and significantly advances our understanding of global dynamics causing ground-based space weather.

\section{Introduction} \label{intro}

The 
interaction of the solar wind and the terrestrial magnetic field produces magnetospheric current systems such as field aligned currents (FACs) which close through the conductive ionosphere, thereby allowing magnetospheric convection to eventuate (e.g. \citealp{Axford1961, Dungey1963, Iijima1976}). For precise investigations of the magnetospheric feedback on the ionosphere and vice versa, an accurate estimate of the ionospheric conductance is critical for realistic global modeling of the magnetosphere, especially during space weather events (e.g. \citealp{Merkine2003}, \citealp{Ridley2004}, \citealp{Merkin2005, Merkin2005a}, \citealp{Liemohn2005}). 
Two dominant sources contribute to the ionosphere's enhanced but finite conductivity - solar extreme ultra-violet (EUV) flux on the dayside, and auroral precipitation in the polar region predominantly on the nightside \citep{schunk_nagy_2009, Newell2009, FullerRowell1987}. Conductance due to solar EUV radiation is relatively well understood through the use of radiative transfer (e.g. \citealp{Chapman1931}). The EUV flux is accounted for in most modern modeling tools as a physics-based empirical function of the solar zenith angle (e.g., \citealt{BREKKE19931493}). Auroral electron and ion precipitation, largely driven by magnetospheric processes, further ionizes neutrals and ions in the ionosphere (e.g., \citealp{Frahm1997, Ahn1998}), and enhances the electrical conductivity in the high-latitude auroral regions \citep{Robinson1987}. 
Since auroral precipitation of charged particles is directly related to variations in the intrinsic magnetic field (e.g., \citealp{Roederer1970}), auroral conductance is an important quantity to predict when investigating the ionosphere's impact on the magnetosphere, and vice versa, during strong driving when the global magnetic field changes rapidly (e.g., \citealp{Welling2019}).

Although several studies have examined the influence of the ionospheric conductance on the global state of the magnetosphere, ionospheric dynamics and their coupled non-linear feedback system
(e.g., \citealp{Raeder2001, Ridley2001, Ridley2004, Liemohn2005, Wiltberger2009, WILTBERGER20041411, Zhang2015, Connor2016, Ozturk2017}), few studies
have actually explored the contribution of conductance on space weather forecasts {{(e.g.} \citealp{Hartinger2017}{)}}, especially during extreme space weather events. This is very difficult to do with data, since measurements of the ionospheric conductance are notoriously inaccurate \citep{Ohtani2014}. Investigations using global models such as \cite{Ridley2004} have indulged in the broad quantification of the conductance due to EUV illumination and auroral precipitation. Studies such as \cite{Wiltberger2009}, \cite{Zhang2015}, \cite{Yu2016} and \cite{Wiltberger2017} addressed this further by identifying the source and impact of various contributors to the auroral conductance. Additional evaluations by \cite{Perlongo2017} 
included the effect of auroral precipitation due to the ring current using a kinetic ring current model coupled to an ionosphere-thermosphere model. Modeling efforts by \cite{Ahn1998}, \cite{Newell2009}, \cite{Korth2014} have estimated ionospheric auroral conductance through empirical relations, using global quantities like solar wind input, ground-based magnetic perturbations and field aligned currents 
as inputs. {{The Robinson conductance model} (\citealp{Robinson1987, Kaeppler2015}) {relating downward precipitating fluxes to auroral conductance is yet another prominent example of empirically-derived conductance from global magnetospheric quantities.}} Recently, \cite{Robinson2018} developed an empirical model using incoherent scatter radar measurements against AMPERE FAC estimations, which spanned the St. Patrick's Day Storm of 2015, an event studied extensively for ionospheric disturbances (e.g., \citealp{Le2016}). In spite of its importance, the impact of auroral conductance during extreme events in global simulations has been hard to determine, due to inaccuracies in conductance estimations within global models, leading to possible underprediction of global quantities like cross polar cap potential (e.g., \citealp{Honkonen2013,  Mukhopadhyay2017}), field aligned currents (\citealp{Anderson2017}), storm indices (\citealp{Liemohn2018b}) and transient ground-based magnetic perturbations \citep{Welling2018}. 

With rising operational usage of first-principles-based geospace models in space weather prediction, the need for accurate conductance models is even more necessary. Operational forecasts of the near-Earth space environment using first-principles based global numerical frameworks (e.g., \citealp{Toth2005}), combining global magnetohydrodynamic (MHD) models (e.g., \citealp{POWELL1999284, Raeder2001}) with suitable inner magnetospheric models (e.g., \citealp{DeZeeuw2004}) and ionospheric models (e.g., \citealp{Ridley2002, WILTBERGER20041411}), have been in use for space weather prediction (\citealp{Liemohn2018}) since the end of the GEM Challenge of 2008-09 (\citealp{Pulkkinen2011, Pulkkinen2013}, Rastaetter et al. 2013). The procedural assessment specifically presented in \cite{Pulkkinen2013} (hereinafter referred to as \emph{Pulkkinen2013}) to investigate predictive skill of global first-principles-based models in predicting ground-based magnetic perturbations $dB/dt$, initiated the transition of model usage toward operational prediction at the NOAA Space Weather Prediction Center (SWPC). 
Several investigations, since then, have further reviewed and systematically addressed the results from this effort, and have suggested rectifications to improve predictive skill (e.g., \citealp{Honkonen2013, Glocer2016, Anderson2017, Mukhopadhyay2017, Liemohn2018, Liemohn2018b, Welling2018}). In particular, the study by \cite{Welling2017} indicated inherent deficiencies in auroral conductance models used in global models that inhibited them from estimating conductance accurately during extreme space weather events. The study concluded that the inability of global models to estimate the ionospheric conductance accurately during extreme events led to underprediction of $dB/dt$.

A key conclusion in the study by \cite{Welling2017} (hereinafter referred to as \textit{Welling2017}) questions the dataset used in estimating a geospace model's auroral conductance during extreme weather, and hypothesizes that the inclusion of information from a larger dataset, including sufficient coverage of extreme events, may lead to improvements in a model's space weather predictive metrics during extreme events. The study falls short of addressing supplementary effects due to the auroral oval's pattern estimation in aforementioned models, and the acute effect such a pattern may have on predictive skill. In this paper, we describe the development and validation of an updated empirical auroral conductance model, 
specifically including data that spans several extreme events, which addresses the concerns raised in \textit{Welling2017}. We use this conductance model within the geospace variant of the Space Weather Modeling Framework (SWMF; \citealp{Toth2005, Toth2012}), identical to the version used operationally at the NOAA Space Weather Prediction Center for space weather forecasting, to investigate the effect of this enhanced conductance model on space weather predictions, and compare these results to the already-existing conductance model within the SWMF. We additionally study the effect of adjusting the pattern of the auroral oval using empirical enhancements based on field aligned current strength, to alter the model's space weather predictions. As a result, in this article, we investigate three major science questions:
 \begin{enumerate}
     \item Addressing \textit{Welling2017}: 
     Does expanding the dataset used to create the initial conductance model help improve space weather predictions?
     \item How significant is the improvement in the space weather predictions due to the enhanced auroral oval adjustment parameters?
     \item Can the combination of the expanded dataset and an auroral oval enhancement cause significant improvement in the global model's space weather prediction? 
 \end{enumerate}
 In order to address the aforementioned questions, a new \textbf{C}onductance \textbf{M}odel for \textbf{E}xtreme \textbf{E}vents (CMEE) has been developed. CMEE is based on the SWMF's empirical auroral conductance model, 
 which 
 uses an inverse-exponential relation to estimate the conductance, and employs an empirically-driven auroral oval adjustment to enhance conductance in regions of strong FACs. A key difference in CMEE, however, is in the dataset it was developed from: CMEE uses one whole year of AMIE data to estimate its conductance. {{Compared to 
 the old model which was derived from the relatively quiet month of January 1997, minute-data from the whole year of 2003 was utilized to develop CMEE. 
 This included some of the most extreme geospace events ever observed} 
 \citep{Cid2015}{.}} In addition to an enlarged training dataset, the value of the empirical coefficients in CMEE are deduced using a non-linear fitting algorithm with suitable extreme boundary conditions that minimizes the absolute error and maximizes the prediction efficiency. 
 %
 %
 The global model configurations used and the science questions addressed in this study, and the subsequent results from this study are described in Sections \ref{methodology} and \ref{result} respectively, while the algorithm used to develop the advance conductance model and the auroral oval adjustment module have been described in Section \ref{cmee}.

\section{Methodology} \label{methodology}


%

\subsection{Simulation Setup}
\label{setup}
\begin{figure}[h!]
	\begin{center}
		\includegraphics[width=0.5\textwidth]{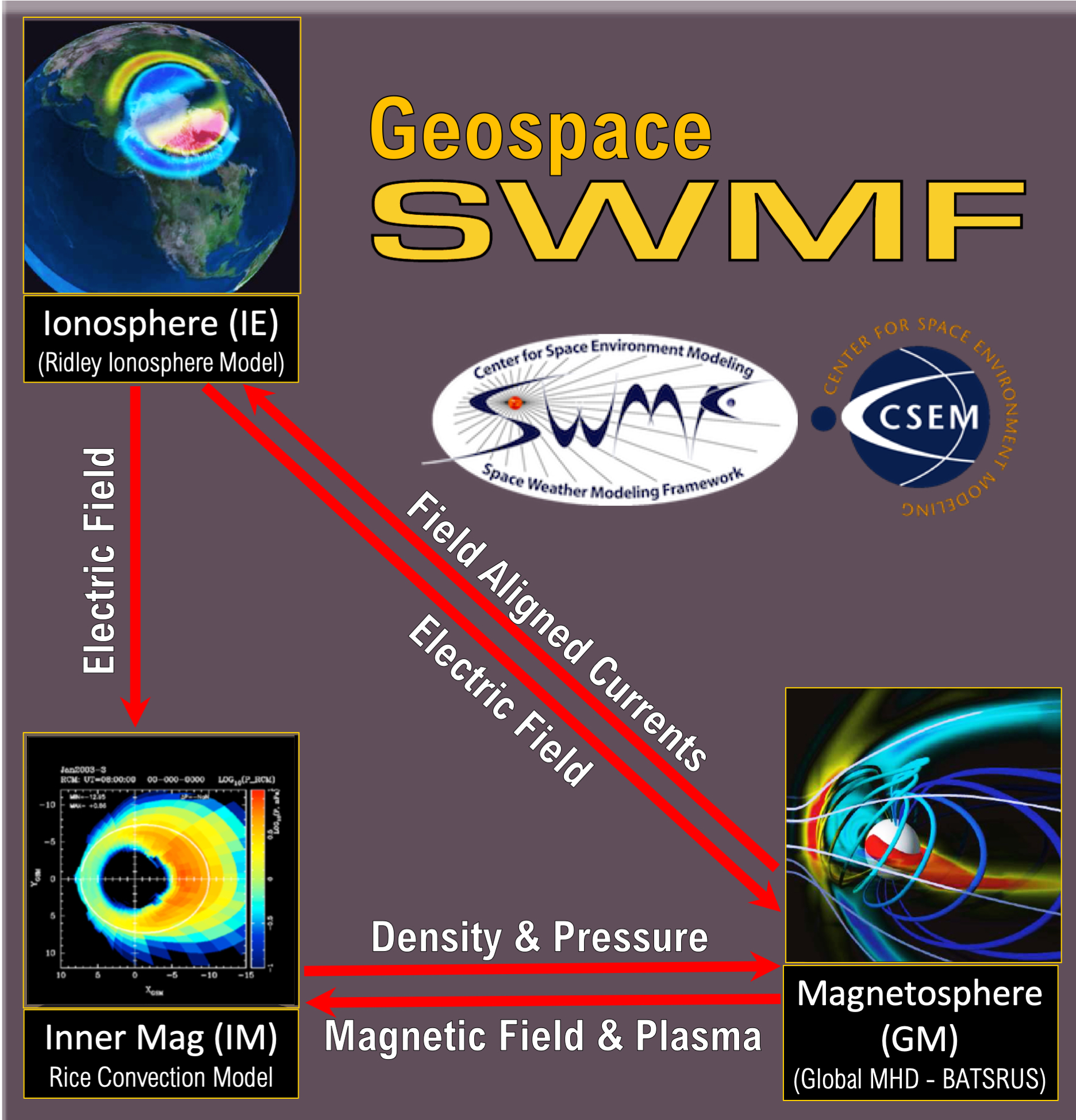}
	\end{center}
	\caption{Component layout of the geospace version of the SWMF, same as the layout in \emph{Pulkkinen2013}, used in this study to investigate the role of auroral conductance in space weather prediction.}
	\label{swmf_setup}
\end{figure}

The SWMF is a flexible framework that executes, synchronizes and couples many otherwise independent models together as one.
It has performed favorably in predictive metric challenges and investigations (e.g., \emph{Pulkkinen2013}; \citealp{Honkonen2013, Mukhopadhyay2017, Welling2017, Liemohn2018b}), contains an easily-modifiable empirical conductance model in the ionospheric electrodynamics module (\citealp{Ridley2004}), and is capable of calculating perturbations to the magnetic field ($\Delta B$) by applying Biot-Savart integrals across its domain to estimate magnetometer values virtually \citep{Yu2010}. For this study, we have used the SWMF with three physical modules activated (Figure \ref{swmf_setup}; details below). Identical to the study conducted by \emph{Pulkkinen2013}, the SWMF's geospace version was configured to use three components: Global Magnetosphere (GM), Inner Magnetosphere (IM), and Ionospheric Electrodynamics (IE). 

The GM module uses the Block Adaptive Tree Solar-Wind Roe Upwind Scheme (BATS-R-US, \citealp{POWELL1999284, Gombosi2003}) model which solves for the ideal non-relativistic magnetohydrodynamic (MHD) equations in the magnetosphere with an inner boundary at $\sim2.5$ Earth radii ($R_E$).
The computational domain for geospace simulations of BATS-R-US extends from $32R_E$ upstream to $–224R_E$ downstream in the $x$ direction and $128R_E$ in the $y$ and $z$ coordinates (GSM). 
The key feature of BATS-R-US is its flexible, block-adaptive Cartesian grid that reserves the highest resolution to regions of interest, ensuring the best combination of performance and accuracy.

The IM region is characterized by closed magnetic field lines and particles of keV energies. This module uses Rice Convection Model (RCM; \citealp{Wolf1982}).
RCM solves for the bounce averaged and isotropic but energy resolved particle distribution of electrons and various ions. RCM receives flux tube volumes from BATS-R-US and returns the pressure and density values to correct those calculated within GM \citep{DeZeeuw2004}. It receives the ionospheric electric potential from the 2-dimensional IE module. 
The density and temperature initial and boundary values are computed from the GM solution.

The IE component calculates height integrated ionospheric quantities at an altitude of about 110 km. To do so, it receives field aligned currents (FACs) from GM and uses the Ridley Ionosphere Model (RIM, \citealt{Ridley2001, Ridley2002, Ridley2004}), a finite-difference Poisson solver, to calculate the electric potential and horizontal currents using a \textit{prescribed} but dynamic conductance pattern.
The module maps FACs at 3.5 Earth radii ($R_E$) over a two dimensional ionospheric domain, solves for the resulting potential using Ohm's Law \citep{Goodman1995}, and returns this value to GM and IM. 
The functioning of and developments to the ionospheric conductance model of RIM are the key features of this article, and are discussed in detail in Section \ref{cmee}, along with the development of a more advanced empirical conductance model, CMEE, as a replacement to the aforementioned model. 


\begin{figure}[h!]
	\begin{center}
		\includegraphics[width=\textwidth]{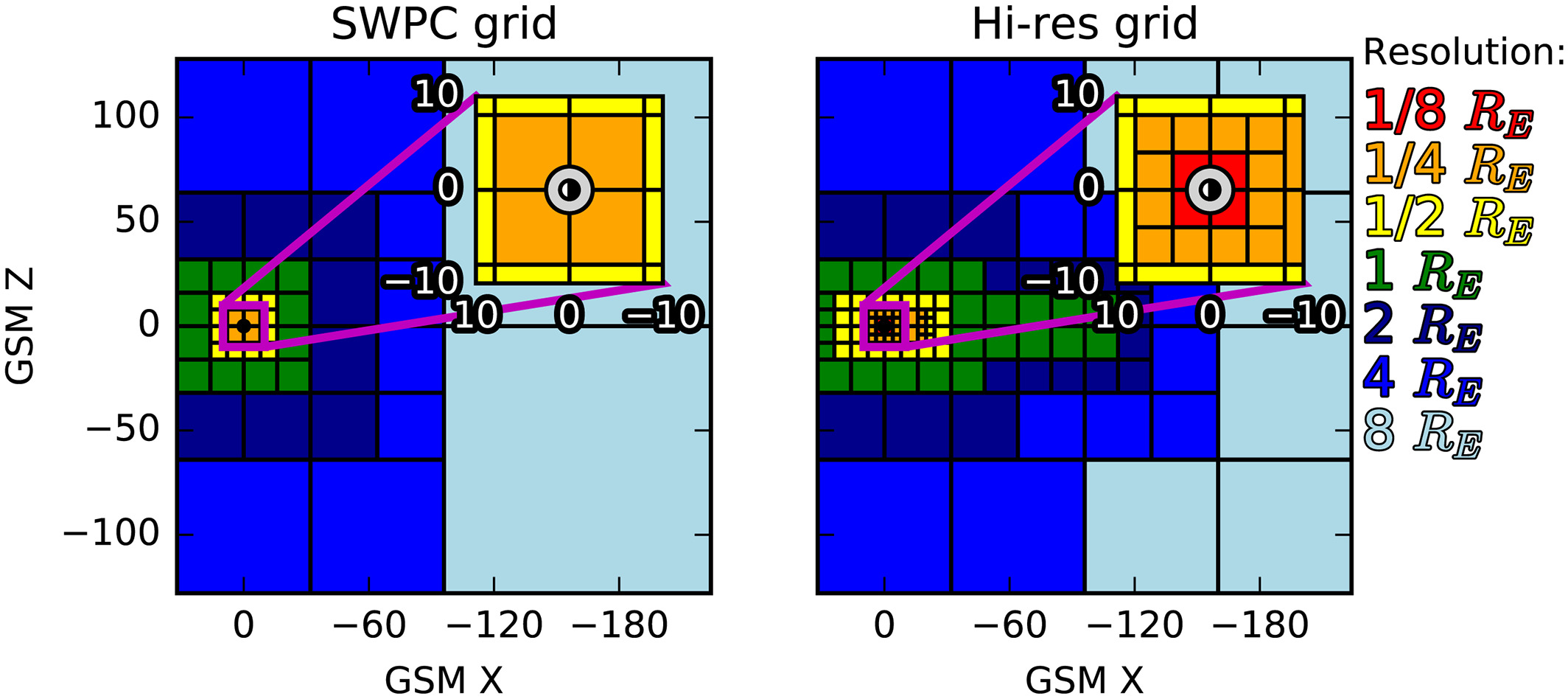}
	\end{center}
    \caption{X-Z cuts showing cell sizes in the two MHD grids (reproduced from \citealp{Haiducek2017}). (Left) The grid used for the \emph{SWPC} configuration (minimum cell size of 0.25 $R_E$). (Right) The higher-resolution grid used for the \emph{Hi-Res SWPC} configuration (minimum cell size of 0.125 $R_E$)}
    \label{mhd_grid}
\end{figure}

In order to simulate a given event, we drive the model using solar wind velocity, magnetic field, density, and temperature, which are used to specify the upstream boundary condition of BATS-R-US. The only other input parameter is F10.7 flux, which is used by IE in computing the dayside EUV-driven ionospheric conductivity (\citealp{Moen1993, Ridley2004}).
Simulation parameters have been kept similar to \emph{Pulkkinen2013}, throughout the study; the model input conditions and parameters are not tailored to individual events. {{The same solar wind values derived in \emph{Pulkkinen2013} from instruments onboard the Advanced Composition Explorer (ACE) satellite were used to drive simulations in the present study.}} For this study, we have simulated the events using two different resolutions of BATS-R-US : \emph{SWPC} and \emph{Hi-Res SWPC} (see Figure \ref{mhd_grid}). The \emph{SWPC} configuration is nearly identical to the \emph{Pulkkinen2013} study, and is used operationally by the Space Weather Prediction Center (SWPC). This grid (Figure \ref{mhd_grid}, left) has cell sizes ranging from 8 $R_E$ in the distant tail to 0.25 $R_E$ at the inner boundary, a 16 $R_E$ diameter cube surrounding the Earth, and contains around 1 million cells. The other configuration, \emph{Hi-Res SWPC}, is similar to the previous configuration but uses a higher-resolution grid (among other modifications), to help resolve field aligned currents at the spatial inner boundary. The cell size of this grid (Figure \ref{mhd_grid}, right) varies from 8 $R_E$ in the tail to 0.125 $R_E$ near the Earth, and contains $\sim$ 1.9 million cells. {{Both configurations use a $91\times181$ cell configuration in the IE domain, with a 2 degree cadence in both latitude and longitude.}} For a detailed description 
of the above configurations, please refer to 
\cite{Welling2010} and 
\cite{Haiducek2017}.


\subsection{Estimation of Auroral Conductance in SWMF} \label{cmee} 

\begin{figure}[t]
	\includegraphics[width=\linewidth]{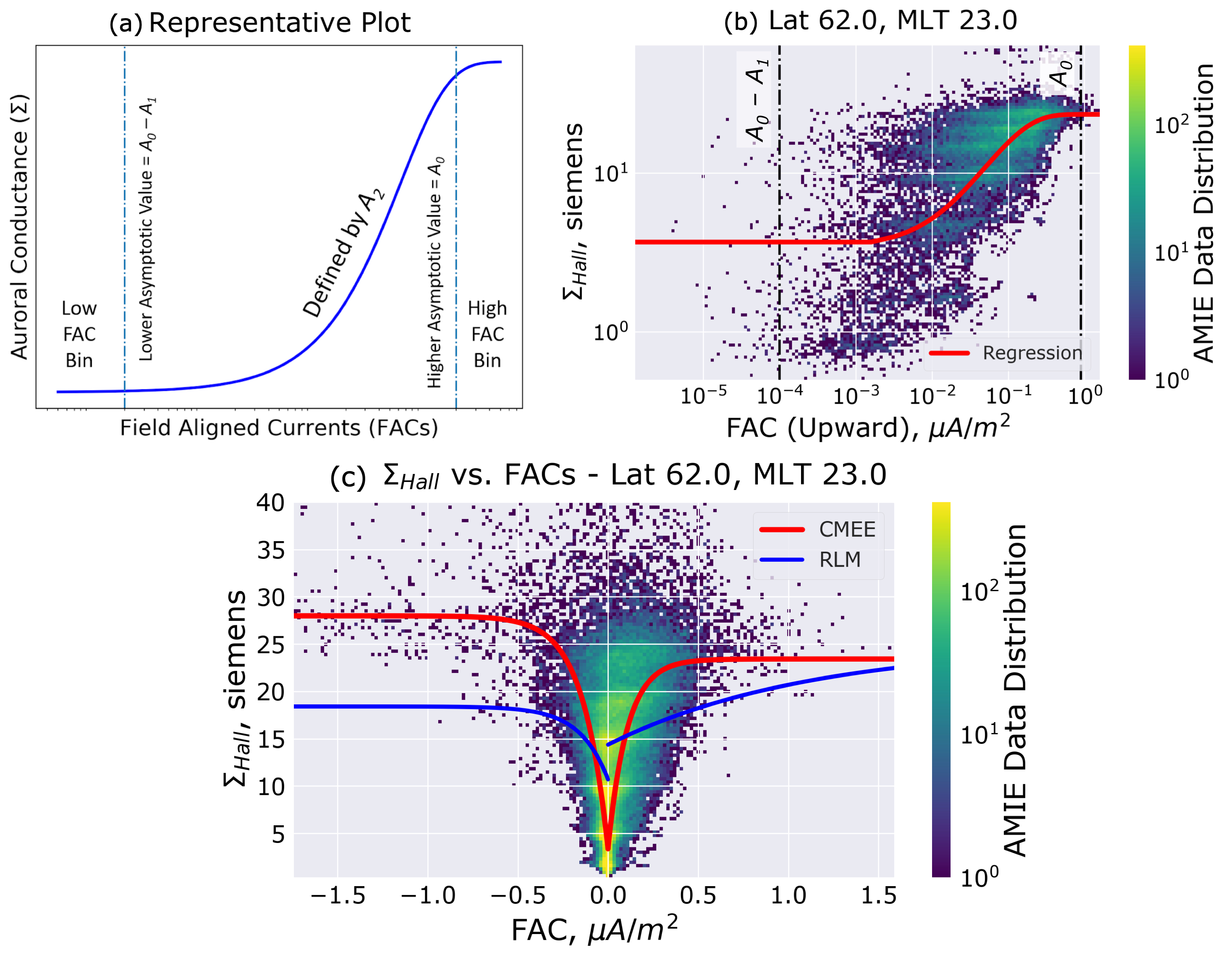}
	\caption{Example Fitting of the Conductance Model for Extreme Events (CMEE) - (a) Representative Line Plot of Auroral Conductance (Hall or Pedersen, in siemens) vs. Field Aligned Currents (FACs, Upward or Downward, in $\mu A/m^2$) through Equation \ref{old_exp} denoting the three regions of interest - low and high FAC bins used to estimate the values of $A_0$ and $A_1$, while the region in between these bins defining the curve using regression of $A_2$. (b) An example log-log plot of the AMIE data showing the scatter of Hall Conductance versus Upward Field Aligned Currents, at magnetic latitude of $68^o$ and magnetic local time (MLT) $23$ in the nightside auroral zone. Alongside the data spread, the regression line is plotted in red with the dot-dashed lines exhibiting the low and high FAC bins. (c) The distribution of AMIE data from 2003 showing the scatter of Hall Conductance versus all Field Aligned Currents plotted along the line plots of RLM and CMEE, denoted in blue and red respectively, at 68$^o$ magnetic latitude and 23 MLT. Note this distribution plot is in linear scale compared to the similar plot part (b), which is in logarithmic scale.}
	\label{curve_fit_demo}
	\centering
\end{figure}



 For Ohm's Law to be solved within IE,
 knowledge of the ionospheric conductance tensor must be known \emph{a priori} (e.g., \citealp{Goodman1995}). Within RIM, the legacy code estimating the ionospheric conductance (\citealp{Ridley2004}) distinguishes two dominant sources of ionospheric conductance: solar EUV conductance on the dayside, and the auroral precipitative conductance in the polar regions. Supplementary sources of conductance, like nightside "starlight" conductance, seasonal dependencies and polar rain, are added as either functions of the dominant sources of conductance, solar zenith angle or scalar constants.
 The solar EUV component to the conductance is dependent on the absorption and ion production function of the atmosphere as a function of the solar zenith angle, and is therefore straightforward to estimate using radiometry; the model described in \cite{Moen1993} is used to estimate this component of the conductance in most global models (e.g. \citealp{Raeder2001} \citealp{WILTBERGER20041411}), including RIM. The conductance due to ion and electron precipitation in the auroral region is harder to predict, as this would require the precise knowledge of the charged particle distribution in the magnetosphere. 
 While a physics-based approach to precipitation has been applied in several global models (e.g. \citealp{Raeder2001}, \citealp{Zhang2015}, \citealp{Yu2016}, \citealp{Perlongo2017}) using kinetic theory (e.g. \citealp{Knight1973}), RIM uses a different and 
 simpler approach to estimate the auroral conductance.
 
 \subsubsection{Functioning of the Ridley Legacy Model} \label{rlm_coeffs}
 
 The auroral conductance module in RIM (briefly described in \citealp{Ridley2004}), hereinafter referred to as the Ridley Legacy Model (RLM), 
 uses the magnitude and direction of {{modelled}} FACs to empirically determine the auroral conductance. This is similar to existing statistical models constructed using FACs to predict and examine precipitation in the auroral ionosphere (e.g. \citealp{Ahn1998}, \citealp{Korth2014}, \citealp{Carter2016}, \citealp{Robinson2018}). While the numerical domain of RIM spans the entire ionosphere, 
 the RLM domain is considerably limited, spanning from the magnetic pole to magnetic latitude of 60$\circ$ for all magnetic local times (MLT). The auroral conductance at a given magnetic latitude and MLT 
 is assumed to have the form: 
\begin{equation} \label{old_exp}
\Sigma_{H or P} = A_0 - A_1 e^{-A_2^2|J_{||}|}
\end{equation}
where $\Sigma_{H or P}$ denotes the auroral Hall or Pedersen Conductance in the ionosphere (in siemens), $J_{||}$ denotes the field aligned current density (in $\mu A/m^2$), and $A_0$, $A_1$ (in siemens) and $A_2$ (in $m/\mu A^{-1/2}$) are fitting coefficients dependent on {{location.}} 
Note that this inverse exponential relation is different from the one mentioned in \cite{Ridley2004}; this was a typographical error and the actual relation 
is given by Equation \ref{old_exp}. 

The empirical coefficients are the result of fitting 
based off of conductance and field-aligned current maps derived from assimilative maps of ionospheric electrodynamics (AMIE; \citealp{Richmond1988, KihnRidley2005}) 
for the month of January 1997 \citep{Boonsiriseth2001}, using ground magnetic perturbations from 
$\sim$150 ground-based magnetometers. AMIE derives the auroral conductance using the formulation in \cite{Ahn1998} and \cite{Lu1997}, which relate ground-based magnetic perturbations to the Hall and Pedersen conductance, and FACs. {{The exact parameters and version of AMIE used in the development of RLM, with further information about the datasets used have been described in detail in} \cite{KihnRidley2005}{.}} 
The month of runs encompasses $\sim 45,000$ two-dimensional maps of Hall and Pedersen conductance and field-aligned currents. 
 In addition to the empirical maps defining the conductance using FACs, additional auroral oval adjustments were applied to constrain and enhance the conductance in regions of strong FAC driving. 


\subsubsection{Conductance Adjustments in the Auroral Oval}
\label{oval_adjust}
The conductance pattern in RLM tends to produce broad regions of 
high conductance that are discontinuous between regions of strong
{{FACs.}}  To improve upon this, an adjustment to the conductance pattern is 
applied to the estimated pattern described above.  The purpose of this is 
to create a channel for electrojets to form in the model and to improve on
the overall electrodynamic result.  Though this feature has been
implemented in RLM for over a decade, this work is the first to 
formally describe it and evaluate its impact.

The algorithm for this adjustment starts by estimating the location of the
auroral oval. {{The location of the oval is updated at each simulation timestep of the ionosphere}}.  Across all local time values ($\phi$) in the model's grid, the
geomagnetic co-latitude of the maximum upward FAC at that local time slice 
($J_{max}(\phi)$) is obtained.  The result is $\theta(\phi)$, or co-latitude 
as a function of local time.
The mean co-latitude, $\theta_{mean}$, weighted by $J_{max}(\phi)$, is then
obtained as follows:
\begin{equation}
    \theta_{mean} = \frac{ \sum
        \theta(\phi) J_{max}(\phi)}{\sum J_{max}(\phi)}
\end{equation}
A day-night shift in the center of the oval is calculated using the 
co-latitudes of $J_{max}(\phi)$ at noon and midnight:
\begin{equation}
    \Delta\theta = \frac{J_{noon} \times (\theta_{noon}-\theta_{mean}) - 
        J_{midnight} \times (\theta_{midnight}-\theta_{mean})}{J_{noon}+J_{midnight}}
\end{equation}
Using these values, the location of the auroral oval is modeled as follows:
\begin{equation}
    \theta({\phi})_{aurora} = \theta_{mean} + \Delta\theta \cos(\phi)
\end{equation}

With the oval location set, an adjustment is applied to the conductance values
about the oval by adjusting the fitting coefficients, $A_0$ and $A_1$:
\begin{equation}
    A_{0,adj} = A_0 e^{- \frac{d^2}{W^2}}
    \label{oval_eq1}
\end{equation}
\begin{equation}
    A_{1,adj} = A_0 - (A_0 - A_1) e^{-\frac{d^2}{W^2}}
    \label{oval_eq2}
\end{equation}
...where, for each line of constant local time, $d$ is the co-latitude distance 
from the oval's locus and $W$ is the width of the oval (default is 
2.5$^{\circ}$).  A baseline conductance about the oval is also applied to
avoid nonphysical solutions in regions of low FACs:
\begin{equation}
    \Sigma_{baseline} = 1.7 \times (\Sigma_{H or P} + ke^{ - \frac{d^2}{W^2}})
    \label{baseline_eq}
\end{equation}
where 1.7 is a multiplier meant to amplify the value of the conductance, and $k$ is a constant derived from the aggregate value of the AMIE-derived auroral conductance in regions of high precipitation (magnetic latitude $\in [65^{\circ},
80^{\circ}]$). {{The 1.7 multiplier is a legacy value and was chosen for robustness and stability of \emph{dB/dt} results.}} In this study, the value of $k$ was found to be 7.5 siemens for 
Hall conductance, and 5 siemens for Pedersen conductance from the AMIE dataset.
The net result of this adjustment is that {{at each timestep}}, about the oval, the range of 
possible conductance values is {{dynamically}} narrowed and enhanced, and a coherent, sharper auroral conductance 
pattern arises.


\subsubsection{Conductance Model for Extreme Events (CMEE)} \label{cmee_coeffs}
 Based on the same formulation as RLM, CMEE was developed using a larger dataset in order to include information during intense space weather events ($Dst < -150 nT$). For this model, minute-resolution data from AMIE for the whole year of 2003 
 were utilized to estimate the new fitting coefficients. {{For consistency, the same version of AMIE} (\citealp{KihnRidley2005}) {used in the development of RLM has been used for the development of CMEE.}} 
 {{The use of a year's worth of minute-data}} significantly increased the model's base dataset from $\sim 45,000$ 2D maps 
 used in RLM, to over $\sim 530,000$ 2D maps used in the present study. {{In addition, the year of 2003 included several intense space weather events. Specifically, the latter half of the year saw some of the largest geomagnetic storms ever recorded by mankind (e.g.} \citealp{Cid2015, Doherty2004} {), while January 1997 (the month off of which RLM is based) hardly saw any event with a \textit{Dst} $\leq$ -100 nT.}} 
In addition to this, 
the value of the empirical coefficients in CMEE are deduced assuming the same empirical relationship between upward or downward FACs with Hall and Pedersen Conductance, as given by Equation \ref{old_exp}. However, unlike RLM which estimates the fitting using equal weighting, the new fitting has been designed using a novel nonlinear regression algorithm which imposes sufficient boundary conditions to ensure that the fitted curve extends to these extreme values and is not just limited to the aggregate value of conductance. This was done by basing the max endpoints of the fittings on the $90\%$ percentile of the FAC values. 


Figure \ref{curve_fit_demo} (a) presents a representative line plot of Equation \ref{old_exp}, and demarcates the conductance vs FAC space into bounded regions designed to estimate fitting coefficients. 
The regression algorithm of CMEE classifies FAC data into low and high magnitude bins, separately for upward and downward FACs. 
The bin boundary for low magnitude FACs, including zero FACs was 
based on 
the approximate order of low magnitude FAC density, where asymptotic behavior of conductance values 
is prevalent and a median value could be found. 
The median value of the conductance populations in this FAC bin is the minima of the curve ($A_0 - A_1$).
For the low FAC case, setting the bin boundary at $\pm 10^{-4} \mu$A/m$^2$ for both upward and downward field aligned currents at all locations led to optimum results. 
%
To deduce the conductance maxima as a constant asymptotic value, 
the FAC dataset was binned into 10 discrete bins with respect to the absolute value of FAC, and the median value of conductance in the bin with the highest FAC values (10th bin) was defined as $A_0$. 
%
A Levenburg-Marquadt (e.g. \citealp{Pujol2007}) type bounded least-squares method was used to estimate the non-linear fitting coefficient $A_2$. 
The fitting error was defined as the arithmetic mean of the median absolute percentage error (MAPE) and the median symmetric efficiency ($\xi$) ratio of the data, as defined in \cite{Morley2018}. 
In order to avoid nonphysical solutions from the ionospheric solver due to large gradients (spikiness) in the conductance values, a smoothing filter was applied on the coefficients. The filter was based on a Laplacian mesh smoothing algorithm (e.g. \citealp{Herrmann1976}), commonly used in image processing \citep{Yagou2002} and mesh refinement \citep{Sorkine2004}. The filter is applied such that at each node $i$,
\begin{equation}
    x_i = 
    \left\{
    \begin{split}
        x_i \quad if \quad \frac{x_i - X}{X} \leq \lambda\\
        X \quad if \quad \frac{x_i - X}{X} > \lambda
    \end{split}
    \right.
\end{equation}
where
\begin{equation}
    X = \frac{1}{N}\displaystyle\sum_{j=1}^{N} x_j
\end{equation}
Here, $\lambda$ is the prescribed difference, $N$ is the number of adjacent vertices to node $i$, $x_j$ is the position of the $j$-th adjacent vertex and $x_i$ is the new position for node $i$. The prescribed difference, similarly defined as the relative difference, is kept at $10\%$.


Figures \ref{curve_fit_demo}(b) shows an example of the fitting using the regression algorithm mentioned above over a map of Hall conductance and FAC distribution from AMIE, at the geomagnetic latitude of 62$^o$ and MLT 23 for upward FACs. Figure \ref{curve_fit_demo}(c) compares the fitting function using CMEE's regression with coefficients from RLM for the same geomagnetic location, but for both upward and downward FACs. 
The usage of a regression algorithm over a larger span of data shows visible differences in Figure \ref{curve_fit_demo}(c), where CMEE, denoted in red, is able to push the max value of the conductance to better estimate the quantity during extreme driving. In addition, because of the usage of low FAC bins, the model is also able to provide uniformity in conductance values when field aligned currents are low and/or switch directions. This was previously not included in RLM, denoted in blue in Figure \ref{curve_fit_demo}(c), as the coefficient values were estimated using uniform weighting on a case-by-case basis separately for upward and downward FACs.



\subsection{Event Selection \& Prediction Assessment} \label{event_pred}

In order to evaluate CMEE's predictive capabilities and address the science questions mentioned in Section \ref{intro}, we have simulated a range of space weather events listed in Table \ref{sims_table}(a) 
using variations of the auroral conductance model within the SWMF 
for comparisons against observations. Since it is a de-facto standard in the space weather community, the present investigation chose to simulate the same events listed in Table 1 of the \emph{Pulkkinen2013} study. 
Simulation of these events was administered for the two resolutions described in Section \ref{setup}, and using four different variations of the conductance model :-
\begin{enumerate}
    \item Using only the empirical coefficients of RLM to specify the aurora,
    \item Using only the empirical coefficients of CMEE to specify the aurora,
    \item \label{point3} Adjusting RLM estimates with the additional enhancements in the auroral oval, and
    \item Adjusting CMEE estimates with the additional enhancements in the auroral oval.
\end{enumerate}
Table \ref{sims_table}(b) lists the 8 sets of simulations resulting from the above combination.

The study uses data from satellite in-situ measurements and ground-based observations for comparisons against model results. Cross polar cap potential (CPCP) from the model variants was compared against values obtained via the AMIE model and observations from the Super Dual Auroral Radar Network (SuperDARN; e.g. \citealp{Khachikjan2008}). 
Since AMIE has a tendency to overpredict CPCP (e.g. \citealp{Gao2012}), observations from the SuperDARN were also used to provide a range to the CPCP estimates. Integrated field aligned currents derived from observations by the Active Magnetosphere and Planetary Electrodynamics Response Experiment (AMPERE) mission \citep{Anderson2014, Waters2020}, estimated using the methodology in \cite{Anderson2017}, were used to compare modeled values of FACs. 
In addition, magnetometer observations from the 12 magnetometer stations listed in Table 2 of the \emph{Pulkkinen2013} study were used 
to evaluate the predicted ground-based magnetic perturbation $\Delta B$ and its temporal variant $dB/dt$.

Using a similar approach as \emph{Pulkkinen2013}, a binary event analysis (e.g. \citealp{Jolliffe2012, Wilks2011}) was used to construct a set of relevant performance metrics. 
An event is defined as the absolute value of a parameter-in-question (any physical quantity like $dB/dt$) exceeding a predetermined event threshold at any time within a comparison window $t_f$. For each such window, four outcomes are possible: "Hit" or True Positive (TP; event is observed, and also predicted), "False Alarm" or False Positive (FP; event is not observed, but predicted by model), "Miss" or False Negative (FN; event is observed, but not predicted), and "Correct No Events" or True Negative (TN; event is not observed, and not predicted). Similar to \emph{Pulkkinen2013}, the analysis forecast window $t_f$ was selected to be 20 minutes. The combined results from all events listed in Table \ref{sims_table}(a) for a given simulation set are divided into discrete events by the forecast window, creating a contingency table accounting for TPs, FPs, FNs and TNs for a specific threshold. Unlike the \emph{Pulkkinen2013} study, 
this study chose to discretize the $dB/dt$ into thresholds ranging from 0.1 nT/s to 1.7 nT/s at intervals of 0.1 nT/s, including the thresholds 0.3 nT/s, 0.7 nT/s, 1.1 nT/s and 1.5 nT/s which were used in the former 
study. In addition to $dB/dt$, the $\Delta B$ values have been discretized using thresholds obtained from \cite{Toth2014} and \emph{Welling2017},  
ranging from 75 to 400 nT at intervals of 25 nT were used. 

Once the contingency tables were prepared for each simulation variation, a combination of performance metrics were applied to study improvements. The metrics used in this study and their respective definitions are listed in Table \ref{metric_tab}. Amongst these metrics, the top four are accuracy measures that help describe the improvement of individual outcomes in a contingency table, while the bottom four metrics quantify the accuracy of a prediction. The Probability of Detection (POD), also called the Positive Prediction Value, is the ratio of positive and negative results, and ranges from 0 to 1, with 1 being a perfect score. 
The Probability of False Detection (POFD) is the ratio of misses against total negative results. 
POFD ranges from 0 to 1, with 0 being a perfect score. Along with the POD, these two ratios are accuracy measures of model discrimination.
The False Alarm Ratio (FAR), also called False Positive Rate is the ratio between the number of negative events wrongly categorized as positive and the total number of actual negative events (false negatives + true negatives).
The Miss Ratio (MR) is defined as the ratio between the number of misses and the sum of hits \& misses, describing the conditional probability of a negative test result given that the condition being looked for is present. Both FAR and MR range from 0 to 1, with 0 being a perfect score. These two metrics are a measure of model reliability. The Threat Score (TS), also known as Critical Success Index is the ratio of all true positives against the sum of total number of occurrences and false alarms. Due to its neglect of non-occurrences, this score is well suited for scoring predictions of rare events like extreme driving during space weather events. The F$_1$ score, another measure of a test's accuracy, 
is defined as the harmonic mean of the POD and the hit rate, given by $(1 - MR)$. Similar to the Threat Score, the F$_1$ score reaches its best value at 1 and worst at 0. 
The True Skill Score (TSS) or Hanssen-Kuiper Skill Score (\citealp{Hanssen1965}) is a performance metric with values ranging from -1 to +1, with 0 representing no skill. The TSS is defined as the difference between the hit rate (given by $1 - MR$) and false alarm rate. Lastly, the Heidke Skill Score (HSS; \citealp{Heidke1926}) is a performance metric that measures the improvements in a model's results against random chance. Similarly to the TSS, the value of HSS 
ranges from -1 to +1, with 0 representing no skill. 
The HSS is popular in space weather forecasting, and has been established as a suitable comparative metric in several space weather studies (\citealp{Welling2010}, \emph{Pulkkinen2013}, \citealp{Toth2014}, \citealp{Welling2018}). 


\begin{table}
    \centering
    \begin{tabular}{c|c}
        \multicolumn{2}{c}{\textbf{(a) List of Events}}\\
        \hline
        \textbf{Event $\#$} & \textbf{Date and Time} \\
        \hline
        1 & 29 October 2003 06:00 UT - 30 October 06:00 UT \\
        2 & 14 December 2006 12:00 UT - 16 December 00:00 UT \\
        3 & 31 August 2001 00:00 UT - 1 September 00:00 UT \\
        4 & 31 August 2005 10:00 UT - 1 September 12:00 UT\\
        5 & 5 April 2010 00:00 UT - 6 April 00:00 UT\\
        6 & 5 August 2011 09:00 UT - 6 August 09:00 UT\\
        \hline
    \end{tabular}
    \begin{tabular}{c|c c c c}
        \multicolumn{5}{c}{\textbf{(b) List of SWMF Simulations}}\\
        \hline
         & \textit{RLM Coeffs} & \textit{CMEE Coeffs} & \textit{RLM w OA} & \textit{CMEE w OA}  \\
        \hline
        \textbf{\emph{SWPC}} & Set A & Set B & Set C & Set D \\
        \hline
        \textbf{\emph{Hi-Res SWPC}} & Set E & Set F & Set G & Set H \\
        \hline
        \multicolumn{5}{l}{\textit{RLM Coeffs}\quad- Empirical Coefficients of the Ridley Legacy Model}\\
		\multicolumn{5}{l}{\textit{CMEE Coeffs}\quad- Empirical Coefficients of the Conductance Model for Extreme Events}\\
		\multicolumn{5}{l}{\textit{RLM w OA}\quad- Ridley Legacy Model, with Auroral Oval Adjustments}\\
	    \multicolumn{5}{l}{\textit{CMEE w OA}\quad- Conductance Model for Extreme Events, with Auroral Oval Adjustments}\\
		\hline
		\hline
    \end{tabular}
    \caption{(a) List of space weather events used in this study to test and validate the different conductance models. This is the same set of events used in \emph{Pulkkinen2013}. (b) A tabular description of all the simulations conducted for this study, binned by SWMF domain variations used: Each set of runs (denoted as 'SET $\times$', where $\times$ is the alphabetic value designated) is a simulation of all space weather events listed in (a), using a particular variation of the auroral conductance model (columns) within a given configuration of the SWMF (rows). }
    \label{sims_table}
\end{table}

\begin{table}
    \centering
    \begin{tabular}{c|c|c}
        \textbf{Performance Metric} & \textbf{Acronym} & \textbf{Mathematical Definition} \\
        \hline
         Probability of Detection & POD & $\frac{TP}{(TP + FP)}$ \\
         Probability of False Detection & POFD & $\frac{FN}{(FN + TN)}$\\
         False Alarm Ratio & FAR & $\frac{FP}{(FP + TN)}$ \\
         Miss Ratio & MR & $\frac{FN}{(TP + FN)}$ \\
         \hline
         Threat Score & TS & $\frac{TP}{(TP + FN + FP)}$ \\
         F$_1$ Score & F$_1$ & $\frac{2TP}{(2TP + FP + FN)}$ \\
         True Skill Score & TSS & $\frac{TP}{TP + FN} - \frac{FP}{FP + TN} = (1 - MR) - FAR$ \\
         Heidke Skill Score & HSS & $\frac{2(TP \times TN - FP \times FN)}{((TP+FP)(FP+TN) + (TP+FN)(FN+TN))}$ \\
         \hline
    \end{tabular}
    \caption{List of performance metrics used in this study.}
    \label{metric_tab}
\end{table}

\section{Results \& Discussion} \label{result} 

\subsection{Impact on Global Quantities}\label{qual_results}

\begin{figure}[h!]
	\begin{center}
		\includegraphics[width=\textwidth]{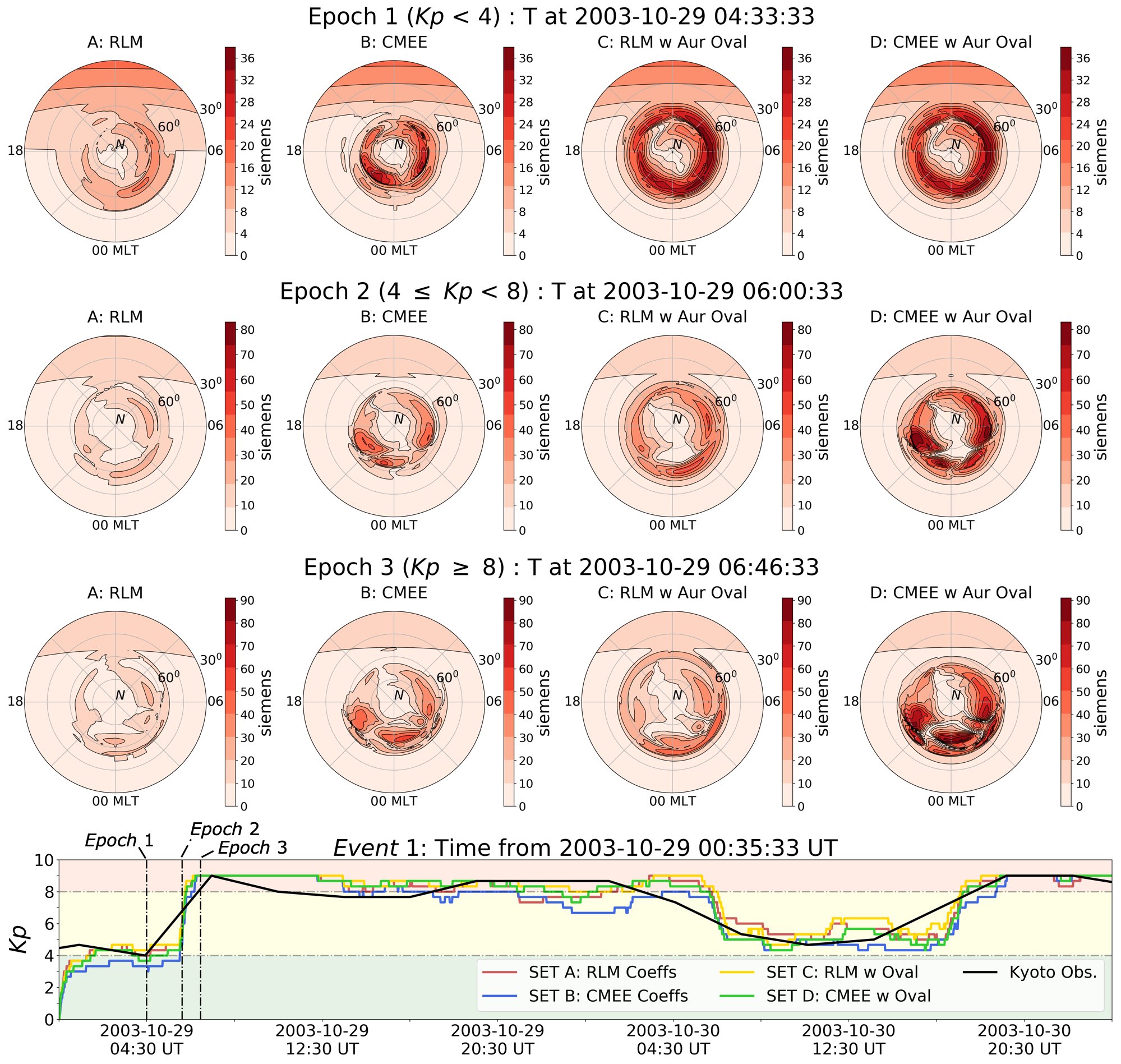}
	\end{center}
    \caption{A comparison of Hall conductance values from different conductance model variants. Dial plots from (left to right) simulation sets A, B, C and D 
    at time instances during Event 1 (Epoch 1, Top Row) when $Kp < 4$, (Epoch 2, Second Row) when $4 \leq Kp < 8$, and (Epoch 3, Third Row) when $Kp \geq 8$. 
    (Bottom Subplot) Comparison of $Kp$ from the Kyoto Observatory (in black) against simulated $Kp$ from simulationsets A (in red), B (in blue), C (in gold) and D (in green). Additionally, the plot background is coloured by the $Kp$, green signifying $Kp < 4$, yellow signifying $4 \leq Kp < 8$, and red signifying $Kp \geq 8$.}
    \label{cond_plot}
\end{figure}

Figure \ref{cond_plot} exhibits the variations in the pattern and magnitude of Hall conductance for simulations using the low-res \emph{SWPC} configuration. 
Each dial-plot column displays the high latitude Hall conductance 
at different time instances from the simulation sets A, B, C and D respectively. 
The first row 
shows results from 04:33 UT on October 29, 2003 : toward the beginning of Event 1, before the sudden commencement 
with the storm index $Kp$ less than 4. 
The second and third rows, titled Epoch 2 and Epoch 3, compare the four sets at 06:20 UT and 06:46 UT on the same day during the sudden commencement and main phase of Event 1, when $4 \leq Kp < 8$ and $Kp \geq 8$ respectively. 
As a reference, the bottom line plot shows the 
$Kp$ throughout the event, along with the predicted $Kp$ from the four simulation variants 
with the background coloured by the magnitude of $Kp$ - green for $Kp < 4$, yellow for $4 \leq Kp < 8$, and red for $Kp \geq 8$. 

Comparing results of Sets A and B, the increased dataset used in CMEE increases the max value of conductance and is capable of 
capturing auroral dynamics across different activity for every epoch. The addition of oval adjustments visibly alters the pattern of conductance - comparison of Sets A and B with their respective counterparts in Sets C \& D illustrate how the adjustments 
intensify the conductance in regions of high field aligned currents, mimicking discrete arcs. The difference in Sets C \& D, while not so apparent in Epochs 1 and 2, are substantially distinct in Epoch 3, when $Kp \geq 8$. In this case, the difference in the conductance caused by the combined usage of the increased dataset spanning extreme events and the additional oval-region enhancement results in a higher conductance peak in Set D. 
For higher $Kp$, 
CMEE increases nightside conductance and lowers dayside conductance.
This is because CMEE coefficients, a byproduct of an increased dataset spanning seasonal changes in addition to being estimated using a nonlinear regression algorithm, computes lower dayside conductance and higher nightside conductance in comparison to the 
RLM coefficients. 
An unusual feature of using FAC-directed empirical models is the emergence of islands of conductance during the peak of the storm (Epoch 3). 
These discontinuities are reduced by the inital usage of the smoothing function on the coefficients, and addition of a baseline value in the auroral oval region. 

\begin{figure}[h!]
	\begin{center}
		\includegraphics[width=\textwidth]{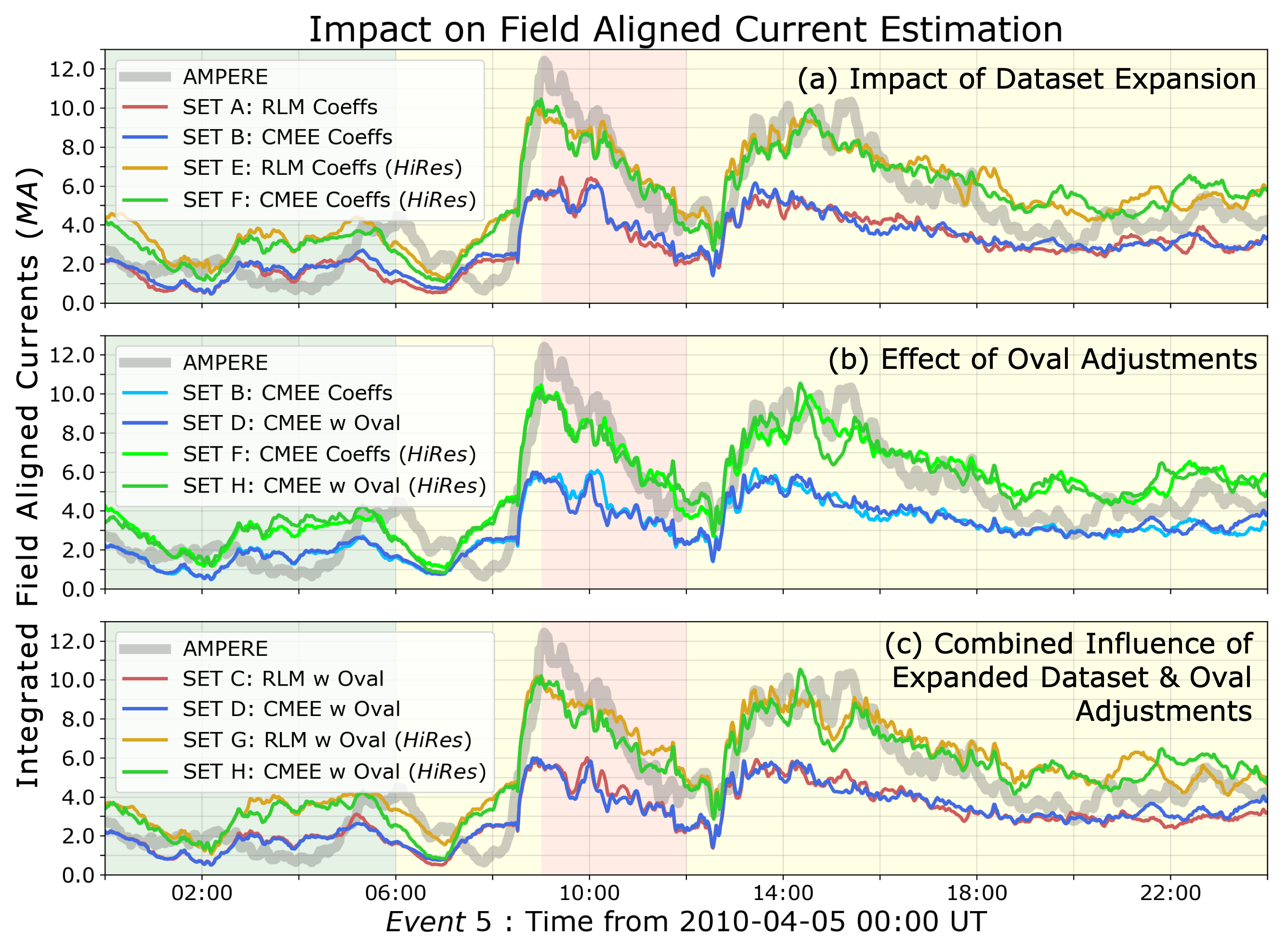}
	\end{center}
    \caption{Time series comparison of integrated field aligned currents (iFACs) for Events 5 spanning the storm main phase from AMPERE (gray line) and the eight simulation sets of the SWMF. Goal of each frame: Top Frame (a) illustrates the impact of dataset expansion on iFACs by comparing Sets A (in red), B (in blue), E (in gold) and D (in green). Middle Frame (b) displays the effect of oval adjustments by comparing Sets B (in light blue), D (in blue), F (in light green) and H (in green). Bottom Frame (c) presents the combined influence of dataset expansion and oval adjustments by comparing Sets C (in red), D (in blue), G (in gold) and H (in green). 
    The plot background is coloured by the $Kp$, green signifying $Kp < 4$, yellow signifying $4 \leq Kp < 8$, and red signifying $Kp \geq 8$.}
    \label{fac_plot_ev5}
\end{figure}

Figure \ref{fac_plot_ev5} compares 
integrated field aligned currents (iFACs) observations during Event 5 by 
AMPERE, 
against estimates from SWMF. 
Events 5 and 6 were observed by AMPERE, and compared to models in \cite{Anderson2017}. 
The iFACs were estimated similarly to \cite{Anderson2017} and were used to compare the effect of dataset expansion in the top panel (a), the impact of oval adjustments in the middle panel (b), and the combined influence both in the bottom panel (c). In each of these panels, we compare the low resolution \emph{SWPC} configuration of the SWMF simulations (Sets A, B, C and D) with the \emph{Hi-Res SWPC} configuration simulations (Sets E, F, G and H) to visualize the impact of conductance on the input conditions to IE. 
While minor variations are caused by the usage of different conductance models, no significant changes are observed either by using the CMEE coefficients or by adjusting the auroral oval. Instead, the results show 
the \emph{Hi-Res SWPC} simulations being able to better capture the magnitude and dynamics of the iFACs than the \emph{SWPC} configurations. {{This is in agreement with results from the study of} \cite{Ridley2010} {who investigated the impact of resolution on ionospheric quantities like FACs, especially with respect to variation in values as we change numerical resolution.}}
While there are definite changes in the FACs and iFAC values due to the different auroral models, the increased resolution helps to capture more of the FACs, dramatically improving the data-model comparison.

\begin{figure}[h!]
	\begin{center}
		\includegraphics[width=0.825\textwidth]{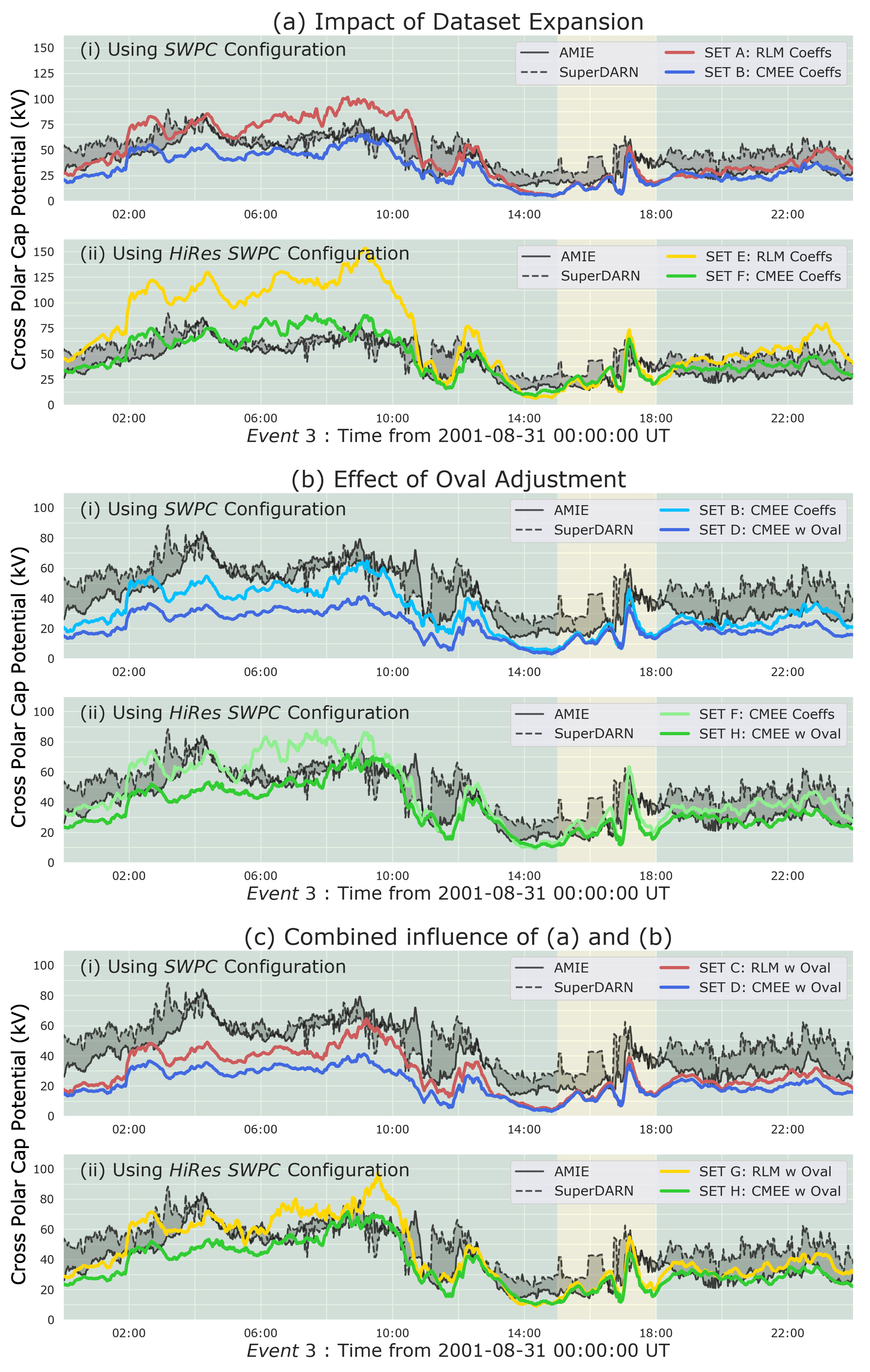}
	\end{center}
    \caption{Time series comparison of cross polar cap potential (CPCP) for Event 3 comparing observations 
    from AMIE, SuperDARN, and the eight configurations of the SWMF. Traces show AMIE in solid black, SuperDARN in dashed black, with the difference region between the datasets coloured gray. The SWMF simulations are coloured similarly to Figure \ref{fac_plot_ev5}. Goal of each frame: Top Frame (a) illustrates the impact of dataset expansion on iFACs by comparing (i) Sets A 
    \& B 
    in upper panel, and (ii) Sets E 
    \& D 
    in bottom panel. Middle Frame (b) displays the effect of oval adjustments by comparing (i) Sets B 
    \& D 
    in upper panel, and (ii) F 
    \& H (in green) in bottom panel. Bottom Frame (c) presents the combined influence of dataset expansion and oval adjustments by comparing (i) Sets C 
    \& D 
    in top panel, and (ii) G 
    \& H 
    in bottom panel. 
    The plot background is coloured by the $Kp$, green signifying $Kp$ $<$ $4$, and yellow signifying $4 \leq Kp < 8$.}
    \label{cpcp_plot}
\end{figure}

Figure \ref{cpcp_plot} compares simulated cross polar cap potential (CPCP) for all simulation sets against values obtained from AMIE and SuperDARN, for Event 3, which was the only event in this study for which high quality AMIE and SuperDARN data were available. 
%
%
Figure \ref{cpcp_plot} is divided into three groups: 
in each group, the low res and high res simulations are compared in separate subplots with the topmost group in part (a) illustrating the impact of updated conductance coefficients on CPCP, middle group in part (b) investigating the impact of oval adjustments, and the bottom group in part (c) comparing the combined influence of dataset expansion and oval adjustments
The difference between the AMIE CPCP, denoted by the solid black line, and SuperDARN CPCP, denoted by the dot-dashed line, has been demarcated using a thick dark grey region in each subplot to give an envelope of expected values based on the observations-based estimates. 

As shown in Figures \ref{cond_plot} and \ref{fac_plot_ev5}, the introduction of CMEE and oval adjustments increases the value of the auroral conductance but does not dramatically impact the strength of FACs, for a given domain resolution. Since the electrostatic potential is the direct output of Ohm's Law, an increment in conductance with no substantial change in FACs leads to a lower value of CPCP. 
This is explicitly observed in part (a), where RLM-driven simulations overestimates the CPCP in both the \emph{SWPC} and \emph{Hi-Res SWPC} cases, in comparison to CMEE-driven simulations. The \emph{Hi-Res} RLM case, denoted in yellow (Frame \ref{cpcp_plot}a-ii), particularly stands out because the FAC-driven conductance reaches the 
ceiling set by the coefficient $A_0$, i.e. as the magnitude of FACs increases, the value of conductance attains the asymptotic maximum value ($A_0$) in the given model. Since the median $A_0$ value is higher in CMEE it is able to give a reasonable CPCP estimate, while RLM's reduced conductance peaks during the strongest driving resulting in the CPCP being an order of magnitude greater. 
In part (b), conductance increments driven by oval adjustments 
largely reduces the CPCP, except during the main phase of the event when $Kp > 4$. This is because, during peak driving, the conductance from both models is so large that the oval adjustments do not affect results substantially. 
In part (c), CMEE-driven CPCP is lower than RLM-driven CPCP, as is expected. The CPCP values from Set D (Frame \ref{cpcp_plot}c-i) are too low, 
indicating that the model is overestimating the conductance which resulted in a lower CPCP. 
For the \emph{Hi-Res} case in Frame \ref{cpcp_plot}c-ii, 
the higher conductance estimation coupled with better resolved FACs acts in favour of CMEE-driven simulations in Set H, and leads to a more realistic CPCP as shown by the comparison against AMIE and SuperDARN. 
In all events, simulations driven with RLM tend to have a higher CPCP compared to CMEE, as the conductance ceiling 
is higher in CMEE than RLM.



\begin{figure}[h!]
	\begin{center}
		\includegraphics[width=0.85\textwidth]{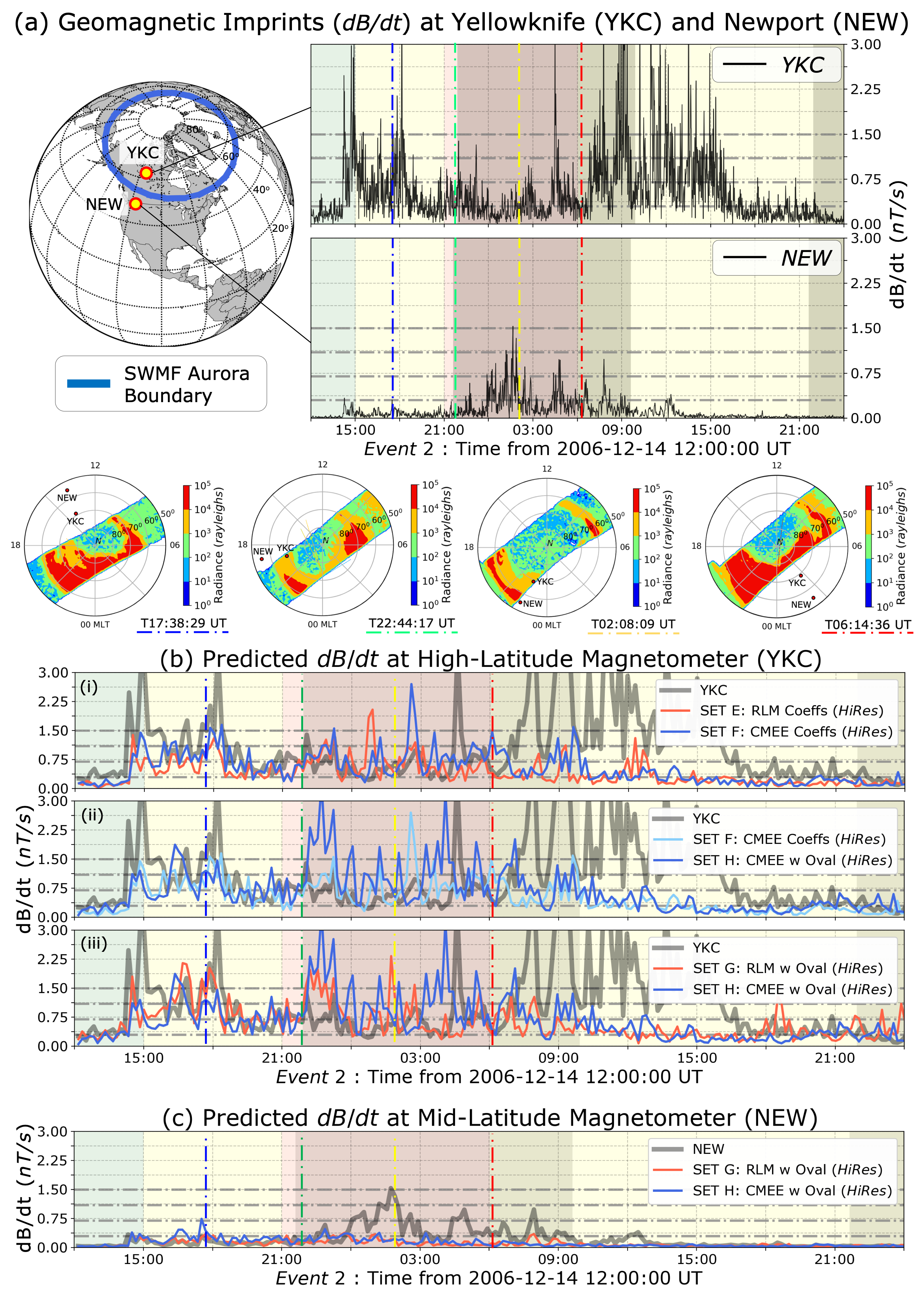}
	\end{center}
    \caption{Impact of changes to the auroral conductance on $dB/dt$ predictions - (a) (Left) Location of Yellowknife (YKC) and Newport (NEW) magnetometer stations mapped in geographic coordinates with the SWMF auroral boundary demarcated using the thick blue line. (Right) Raw $dB/dt$ observations at a 1-minute cadence at YKC and NEW. 
    {{(Bottom) Expansion of the auroral oval as seen through DMSP F16 auroral radiance maps and the magnetometer stations at Yellowknife (YKC) and Newport (NEW). The dialplots on top are demarcated by blue, green, yellow and red dot-dashed lines in the line plots, in increasing order of their timestamps.}}
    (b) Comparison of max-filtered predicted $dB/dt$ from \emph{Hi-Res} SWMF simulations against similarly filtered $dB/dt$ observations at Yellowknife (YKC). Goal of each panel: Top panel (i) shows impact of coefficients by comparing simulation sets E (in red) and F (in blue). Middle panel (ii) illustrates the impact of oval adjustments by comparing sets F (in light blue) and H (in blue). Bottom panel (iii) compares sets G (in red) and H (in blue). Observations are shown as a thick, grey curve. (c) Comparison of max-filtered predicted $dB/dt$ from sets G (in red) and H (in blue) against observations (thick, grey curve). The dot-dashed lines in the line plots are markers of the thresholds used in the \emph{Pulkkinen2013} study for their event-based analysis. The background of the line plots are coloured by $Kp$, similarly to Figure \ref{fac_plot_ev5}. The dark shaded background regions are times when the respective magnetometer was in the nightside.}
    \label{dbdt_fig}
\end{figure}

Figure \ref{dbdt_fig} illustrates the impact of conductance on $dB/dt$ predictions during Event 2, 
at two magnetometer stations - the high-latitude magnetometer station at Yellowknife (YKC) located at magnetic latitude (MLat) $68.93^\circ$ N and magnetic longitude (MLon) $299.36^\circ$, and the mid-latitude magnetometer station at Newport (NEW) located at MLat $54.85^\circ$ N and MLon $304.68^\circ$. While YKC and NEW are far apart latitudinally, longitudinally they are separated by less than $5^\circ$, making them a good candidate to study the expansion of the auroral oval under strong driving conditions. The background in each subplot, in addition to being coloured by $Kp$ similar to Figures \ref{fac_plot_ev5} and \ref{cpcp_plot}, are darkened to indicate times when the magnetometer was on the nightside. Additionally, dash-dot lines in all subplots indicate the four thresholds chosen in the \emph{Pulkkinen2013} study.

Between 14:08 UT and 18:17 UT on December 14, 2006, as activity increases, massive $dB/dt$ spikes were observed at YKC with values 
crossing the four \emph{Pulkkinen2013} thresholds. These spikes died down as activity increased, indicated by the increment in the $Kp$ values. From $\sim$18:20 UT to 07:04 UT on December 15, except for one massive spike at 04:28 UT, $dB/dt$ spikes at YKC barely cross the second and third threshold. During this time period, the magnetometer was mostly on the nightside. Interestingly, all substantial perturbations observed  
at NEW occur during this same time interval, between 22:21 UT and 07:54 UT. This is an indication that the auroral oval expanded equatorward 
during this given time interval {{as shown by the auroral radiance measurements by Defence Meteorological Satellite Program (DMSP) F16 passes}}, with the storm intensifying. 
This expansion of the oval resulted in latitudinally-high YKC no longer being in the auroral zone and instead being in the 
polar cap region, while the lower boundaries of the auroral oval reached latitudinally-lower NEW. 
Starting at 07:54 UT, spikes at NEW died down and were almost negligible throughout the rest of the event. Around the same time, massive spikes crossing all four thresholds were observed again at YKC as the magnetometer station approaches the midnight-dawn sector. The spikes at YKC were observed until 16:33 UT as the magnetometer station rotated to the dawn-noon sector, through the recovery period of the event. 

In parts (b) and (c) of Figure \ref{dbdt_fig}, modeled $dB/dt$ at YKC and NEW are compared against observations. The topmost panel in part (b) compares modeled $dB/dt$ from Sets E and F 
addressing the impact of dataset expansion. The middle panel in (b) compares Sets F and H 
to address the effect of auroral oval adjustments, while the bottom panel compares Sets G and H 
to study the combined influence of both the expanded dataset and the oval adjustments. In part (c), modeled $dB/dt$ from Sets G and H are compared against observations at NEW. 
To simplify visualization, the minute-resolution data from both observed and modeled $dB/dt$ values in parts (b) and (c) 
have been max-filtered for every 10 minute interval. Additionally, the subplot background and threshold lines in parts (b) and (c) are plotted and coloured similarly to part (a).

In the top panel of part (b), 
the magnitude of the CMEE-simulated $dB/dt$ spikes are mostly at par with or moderately larger than the RLM-simulated spikes through most of the event. Both Sets E and F reasonably modeled the $dB/dt$ during the time interval when the oval expanded and YKC was in the polar cap. However, they were unable to reproduce the heavy spikes that appeared both before and after the time interval, barely crossing the fourth threshold of 1.5 $nT/s$ at any given instance. In the middle panel, both the frequency and magnitude of the $dB/dt$ spikes increased significantly with the introduction of the oval adjustments. 
{\color{black}While this led to minor improvements in reproducing observations at time intervals when YKC observed heavy spikes, a substantial change occured during the oval expansion when there were minimal $dB/dt$ perturbations in both the observations and the coefficient-driven simulation results but intense spikes at high frequencies in the oval-adjusted simulation output.}
This increment in $dB/dt$ spikes is dominant in the bottom panel of part (b) in both CMEE and RLM driven simulations. The impact of the dataset expansion combined with the oval adjustment in Set H simulations led to a sharp increase in the magnitude of the spikes, in addition to the sharp rise in frequency. 
Part (c) indicate that the model does not reproduce the $dB/dt$ spikes at NEW, regardless of the conductance model used. This is in direct contrast to the results from the last panel of part (b) which compares the same model variants but shows multiple intense $dB/dt$ spikes at YKC during the same time interval. This indicates that while usage of CMEE + oval adjustments improved the performance, there were still outstanding issues concerning the expansion and location of the oval that may require a more comprehensive, physics-based approach. 

\begin{figure}[h!]
	\begin{center}
		\includegraphics[width=0.8\textwidth]{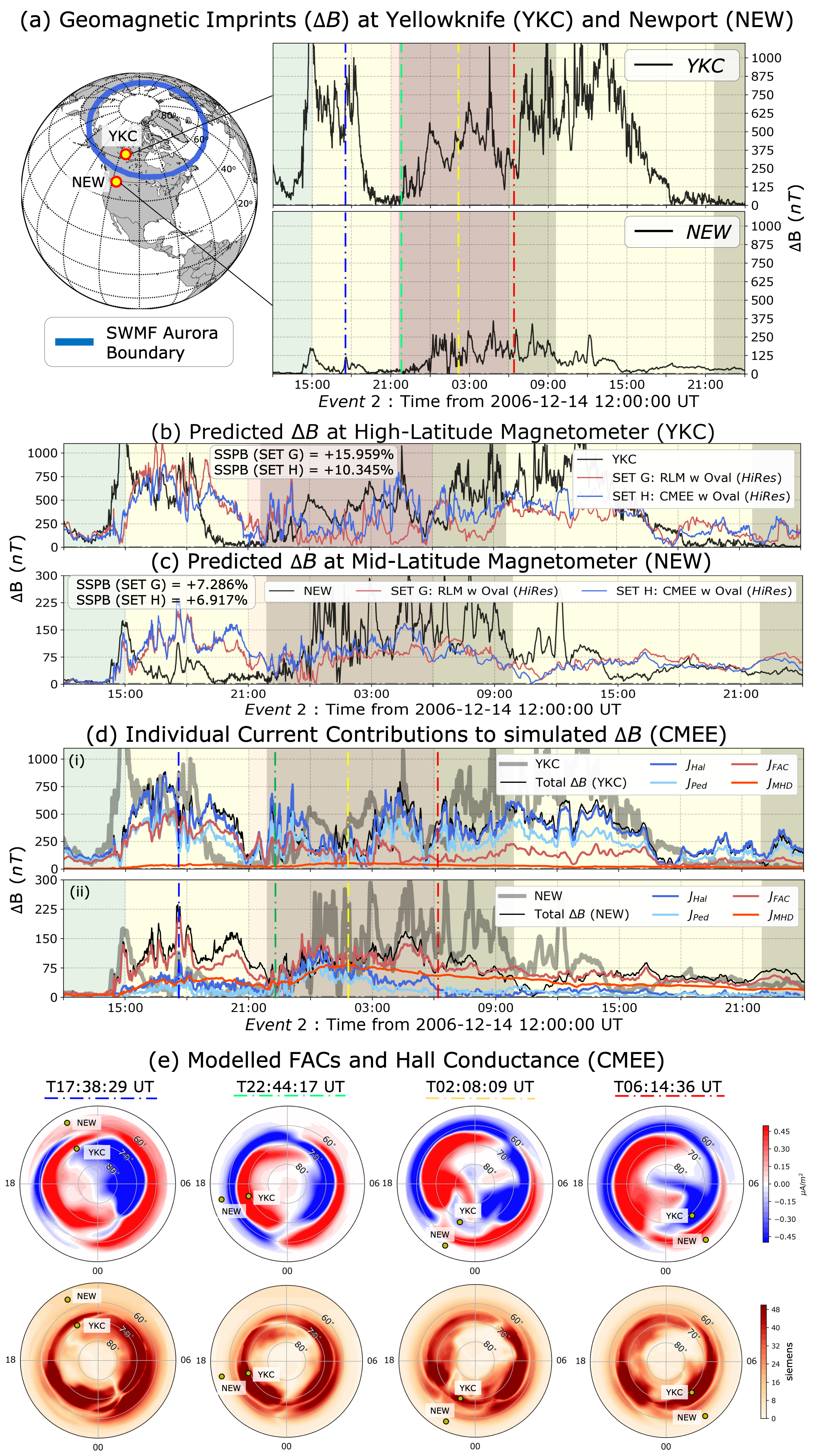}
	\end{center}
    \caption{Impact of changes to the auroral conductance on $\Delta B$ predictions - (a) (Left) Location of Yellowknife (YKC) and Newport (NEW) magnetometer stations mapped in geographic coordinates with the SWMF auroral boundary demarcated using the thick blue line. (Right) Raw $\Delta B$ observations at a 1-minute cadence at YKC and NEW. (b) Comparison of predicted $\Delta B$ from \emph{Hi-Res} SWMF simulations against observations at YKC, and (c) at NEW. Both subplots compare results from simulation sets G (in red) and H (in blue) against observations (in black). (d) Comparing contribution of individual current sources in the simulated $\Delta B$ at (i) YKC and (ii) NEW. The contributions from Hall currents are in blue, Pedersen currents in light blue, FACs in red, and MHD in orange. The background of the line plots are coloured by $Kp$, similarly to Figure \ref{fac_plot_ev5}. The dark shaded background regions are times when the respective magnetometer was in the nightside. {{(e) Dial plots of modelled FACs (top row) and Hall Conductance (bottom row) in the Northern hemisphere from simulation set H at the same time instances as the DMSP passes in Figure \ref{dbdt_fig}.}} }
    \label{delb_fig}
\end{figure}

Figure \ref{delb_fig} illustrates comparison magnetic perturbations $\Delta B$ at the same magnetometer stations during the same event to provide further clarity on the issue of auroral expansion. 
Part (a) 
compares the modeled and simulated $\Delta B$ at YKC and NEW during the event. 
At YKC, 
heavy fluctuations were observed in the $\Delta B$ values 
corresponding with the same time intervals when the massive spikes in $dB/dt$ were observed in Figure \ref{dbdt_fig}(a): between 14:21 UT and 18:19 UT, on December 14, and 06:42 UT and 17:07 UT on December 15. The magnitude of $\Delta B$ were $\geq$ 500 $nT$ during these time intervals. 
At NEW, while all variations in $\Delta B$ were comparatively lower ($\leq$ 400 $nT$), heavy fluctuations 
were seen during the same time interval when the auroral oval expands and significant $dB/dt$ perturbations in Figure \ref{dbdt_fig}(a) occur, between 23:37 UT and 12:07 UT. 
During the oval expansion phase, 
YKC-observed $\Delta B$ increases steadily with time producing {\color{black}minimal fluctuations} during this period, retroactively indicating why the $dB/dt$ is low.


In parts (b) and (c) of Figure \ref{delb_fig}, the simulated $\Delta B$ from Sets G and H 
reasonably reproduce the observed $\Delta B$ pattern. 
During the oval expansion phase of the event, the simulated $\Delta B$ of both sets fluctuate with higher frequency and magnitude than is observed at YKC, 
thereby explaining the massive spikes in the simulated $dB/dt$ 
seen during the same time interval 
in Figure \ref{dbdt_fig}(b). 
Quantitatively, the Set H simulations 
exhibit the best performance with a symmetric signed bias percentage (SSPB; \citealp{Morley2018}) of $\sim 5.6\%$. Here, SSPB measures the symmetric bias in the forecast against the observed values. 
At NEW, comparison of the simulated $\Delta B$ from either sets do not differ substantially with each other, 
with a negligible difference of $\leq 1\%$ in their respective SSPB. Neither models are able to predict the perturbations during the main phase of the storm between 00:00 UT to 09:00 UT, explaining similarly poor performance in predicting the $dB/dt$ values for this magnetometer. 
Part (d) compares the individual contributions of the global current systems - auroral Hall and Pedersen currents, field-aligned currents and magnetospheric currents, in the $\Delta B$ estimation at YKC and NEW from the Set H simulation. At YKC, auroral and field-aligned currents are the dominant current systems driving perturbations in the magnetic field while magnetospheric currents contribute negligibly.
The opposite is true at NEW, 
where the $\Delta B$ variations are mostly driven by changes in the magnetospheric currents and field aligned currents, with auroral currents barely affecting the simulated $\Delta B$ even during the peak driving of the system, indicating minimal contribution. {{This is further corroborated by the dial plots in Part (e) with the top row showing the extent of saturated field aligned currents in the SWMF domain and compares it to the domain boundary of the modeled auroral conductance in the bottom row which clearly halts at 60 degree MLat.}}

The comparisons in Figures \ref{dbdt_fig} and \ref{delb_fig} indicate that in the modeled $\Delta B$ and $dB/dt$ values, the auroral currents have little or no impact on mid and low latitude magnetometer predictions as the auroral oval is not able to extend equatorward to these latitudes. While this is expected during quiet conditions, the impact of auroral currents during extreme events can change dynamically with the expansion of the auroral oval, and can extend to much lower latitudes as evidenced by NEW during this event. 
The impact of this shortcoming on predictive skill has been described in further detail in Section 
\ref{analysis}.

\subsection{Performance Quantification of $dB/dt$ Comparisons}
\label{dBdt_metric_section}
The results from the binary event analysis performed on the $dB/dt$ predictions show that changing the auroral conductance in the global model, either by expanding the dataset or by applying the oval adjustments, 
led to minimal or no improvement in skill score for the lowest $dB/dt$ threshold, but improved skill for the remaining $dB/dt$ thresholds, with the most 
improvement in the highest thresholds.
Table \ref{hss_dbdt_table} presents a re-analysis of the results from \emph{Pulkkinen2013}, emphasizing the changes in the HSS of $dB/dt$ results, that were caused by CMEE and the auroral oval adjustments. 
In part (a) of the table, the expansion of dataset results in 
the improvement of HSS in each threshold for both the low and high resolution cases, as evidenced by the difference column. 
This addresses \emph{Welling2017}'s original question, that expansion of the dataset can lead to improvement in $dB/dt$ predictions. 
In part (b), 
the HSS improvement caused by oval adjustments to the aurora is more substantial than in part (a), with HSS going up by $\sim 0.1$ in the highest thresholds for both \emph{SWPC} and \emph{Hi-Res SWPC} configurations. The comparison of both RLM and CMEE combined with oval adjustments in case (c) show similar improvements in predictive skill for the higher $dB/dt$ thresholds when using CMEE with oval adjustments. 

\begin{figure}[h!]
	\begin{center}
		\includegraphics[width=\textwidth]{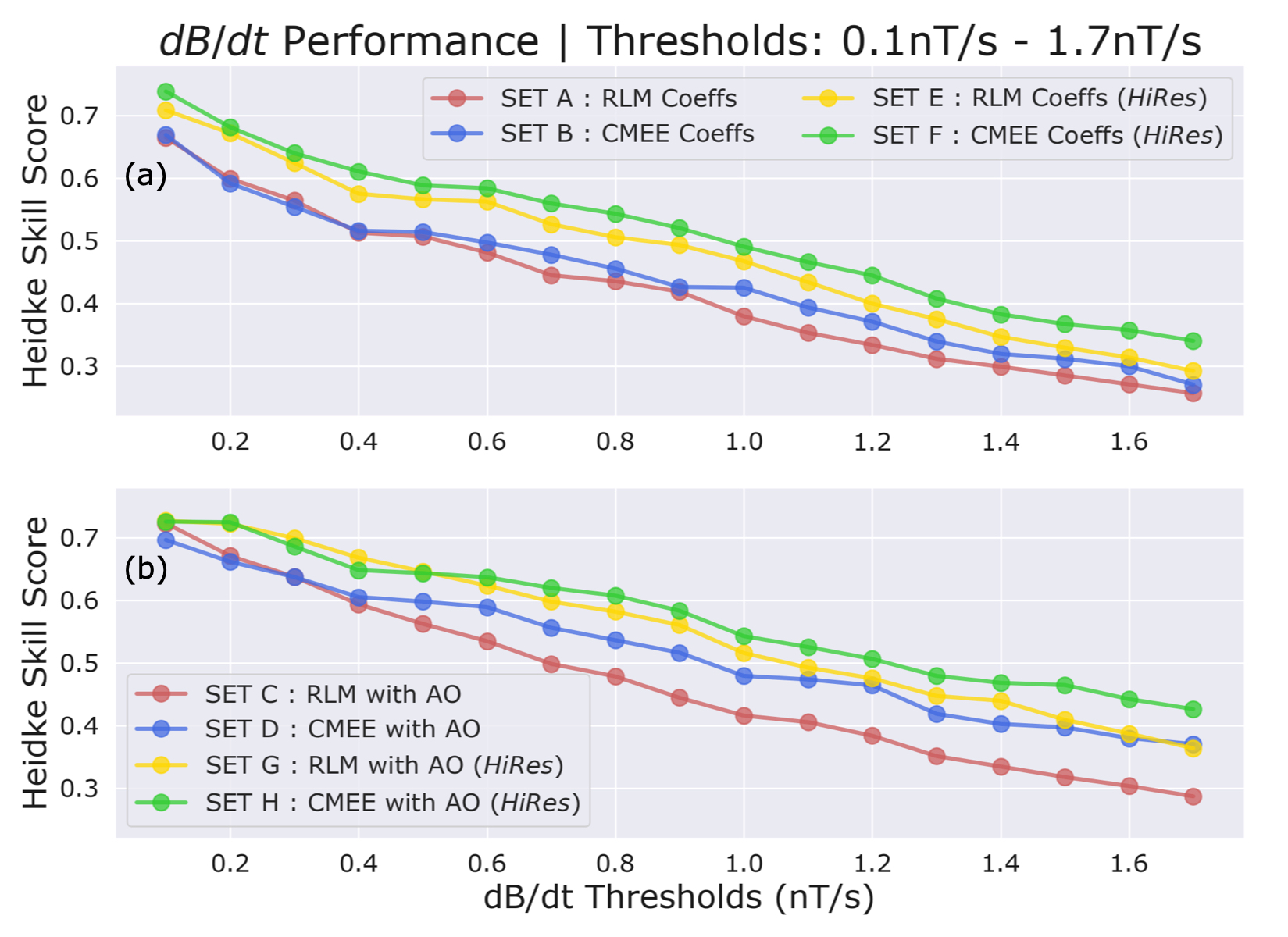}
	\end{center}
    \caption{Heidke Skill Score (HSS) Performance of all SWMF simulation variants at ascending $dB/dt$ predictions for all events from Table \ref{sims_table}(a). (a) Comparison of simulation sets A (in red), B (in blue), E (in yellow) and F (in green) illustrating the impact of dataset expansion. 
    (b) Comparison of simulation sets C (in red), D (in blue), G (in yellow) and H (in green) displaying the overall impact of dataset expansion with oval adjustments. 
    Note the y-axis in (a) and (c) does not start at zero.}
    \label{hss_dbdt_plot}
\end{figure}

Figures \ref{hss_dbdt_plot}(a) and (b) provide a quantitative picture of HSS improvement in the $dB/dt$ predictions over many more 
thresholds. 
In both subplots, the $y$-axis is HSS, while the increasing $dB/dt$ thresholds on the $x$-axis provide a quantitative value of space weather activity. As expected, the HSS scores for all models decreased with increasing threshold value. However, in the most-extreme thresholds CMEE-driven simulations out-peform RLM-driven simulations, 
with improvements in the HSS of the same order as previously evidenced in Table \ref{hss_dbdt_table}. 
%
%
The HSS values in the highest dB/dt thresholds for the low-resolution runs of CMEE, in both parts (a) and (b), were either at par or larger than the HSS values for not only the low-resolution but also the high-resolution RLM simulations. This is a significant improvement in the skill score due to CMEE, as this provides an alternate physics-based remedy that otherwise could only be solved numerically.
Naturally, the HSS values of the high-resolution CMEE-driven simulations were the highest at almost all thresholds.
Using this result, we can partially address the science questions posed in Section \ref{intro} that the auroral conductance impacts the $dB/dt$ significantly, and that improvements in the magnitude or pattern of the conductance boosts predictive skill scores for strong driving of the system.


To better 
quantify the variation in model performance, 
the values of all performance metrics listed in Table \ref{metric_tab} were investigated. 
Table \ref{dbdt_metric_table} presents these metrics 
calculated for all model variants at the high dB/dt threshold of $1.5$ nT/s. 
In this table, the results show the \emph{SWPC} configuration in the left and the \emph{Hi-Res SWPC} configuration in the right, with the worst performance by configuration coloured in orange and the best performance coloured in blue. 
For both the \emph{SWPC} and \emph{Hi-Res SWPC} configurations, 
the POD and MR improved quite significantly for CMEE and the oval adjustments, indicating that the number of hits and misses 
increased and decreased, respectively. In addition, all skill score metrics in the latter half of the table, excluding TSS, indicate best performance for CMEE with oval adjustment variant for both resolutions of the model. The TS and F$_1$ score increased indicating that the number of hits increased. As has been shown in the previous figure and table, 
the HSS improves as we switch models to introduce oval adjustments and expansion of the dataset. 
However, the opposite occured 
when looking at POFD and FAR values were considered: the application of oval adjustments led to sharply increased FAR values in both low and high res configurations. 
While the hits and true negatives increased significantly and misses decreased, as supported by the POD and MR values, the number of false alarms increased steadily as the conductance coefficients were changed and jumped significantly with the application of the oval adjustments. This indirectly affected the TSS, which is defined as the difference between the hit rate and miss rate, or mathematically as 1 - (FAR + MR). Since the FAR increased, in spite of the decreased MR, TSS values reduced by more than 0.05 as we switched models. Given that this order of change in skill was similar to what was achievable by changing model resolutions, the increment in false alarms is a significant drawback when using oval adjustments. The aforementioned trend was observed in all $dB/dt$ thresholds from 0.7 nT/s and above, indicating that this was not an isolated case. The performance metrics for the other thresholds have been presented in the supp. material. 

\begin{table}[h!]
	\begin{center}
        \begin{tabular}{p{0.7in}|| p{0.5in} p{0.5in} p{0.6in} | p{0.5in} p{0.5in} p{0.6in}}
		    \multicolumn{7}{c}{\textbf{(a) Impact of Dataset Expansion}}\\
		    \hline
		    \multirow{2}{*}{Threshold}& \multicolumn{3}{c}{\emph{SWPC} Configuration} & \multicolumn{3}{c}{\emph{Hi-Res SWPC} Configuration}\\
		    & RLM & CMEE & Difference & RLM & CMEE & Difference\\
		    \hline
		    \textbf{0.3 nT/s} & 0.521
		    & 0.554 & \textcolor{teal}{$+0.033$} & 0.624 & 0.640 &  \textcolor{teal}{+0.016}\\
		    \textbf{0.7 nT/s} & 0.445 & 0.478 & \textcolor{teal}{+0.033} & 0.526 & 
		    0.559 &  \textcolor{teal}{+0.033}\\
		    \textbf{1.1 nT/s} & 0.353 & 0.394 & \textcolor{teal}{+0.040} & 0.434 & 0.466 & \textcolor{teal}{+0.032}\\
		    \textbf{1.5 nT/s} & 0.285 & 0.312 & \textcolor{teal}{+0.027} & 0.330 & 0.367 & \textcolor{teal}{+0.037}\\
		    \hline
		    \multicolumn{7}{c}{\textbf{(b) Effect of Oval Adjustment (OA)}}\\
		    \hline
		    \multirow{2}{*}{Threshold}& \multicolumn{3}{c}{\emph{SWPC} Configuration} & \multicolumn{3}{c}{\emph{Hi-Res SWPC} Configuration}\\
		    & CMEE & CMEE$^+$ & Difference & 
		    CMEE & CMEE$^+$ & Difference\\
		    \hline
		    \textbf{0.3 nT/s} & 0.554 & 0.637 & \textcolor{teal}{+0.083} & 0.640 & 0.685 & \textcolor{teal}{+0.046}\\
		    \textbf{0.7 nT/s} & 0.478 & 0.556 & \textcolor{teal}{+0.078} & 0.559 & 0.619 & \textcolor{teal}{+0.060}\\
		    \textbf{1.1 nT/s} & 0.394 & 0.474 & \textcolor{teal}{+0.080} & 0.466 & 0.525 & \textcolor{teal}{+0.059}\\
		    \textbf{1.5 nT/s} & 0.312 & 0.397 & \textcolor{teal}{+0.085} & 0.367 & 0.465 & \textcolor{teal}{+0.098}\\
		    \hline
		    \multicolumn{7}{c}{\textbf{(c) Influence of Dataset expansion and OA Combination}}\\
		    \hline
		    \multirow{2}{*}{Threshold}& \multicolumn{3}{c}{\emph{SWPC} Configuration} & \multicolumn{3}{c}{\emph{Hi-Res SWPC} Configuration}\\
		     & RLM$^+$ & CMEE$^+$ & Difference & RLM$^+$ & CMEE$^+$ & Difference\\
		     \hline
		    \textbf{0.3 nT/s} & 0.637 & 0.637 & $\pm$0.000 
		    & 0.699 & 0.685 &  \textcolor{red}{$-0.013$}\\
		    \textbf{0.7 nT/s} & 0.498 & 0.556 & \textcolor{teal}{+0.058} & 0.598 & 0.619 & \textcolor{teal}{+0.022}\\
		    \textbf{1.1 nT/s} & 0.406 & 0.474 & \textcolor{teal}{+0.068} & 0.492 & 0.525 &   \textcolor{teal}{+0.033}\\
		    \textbf{1.5 nT/s} & 0.318 & 0.397 & \textcolor{teal}{+0.079} & 0.409 & 0.465 &   \textcolor{teal}{+0.056}\\
		    \hline 
		    \multicolumn{7}{l}{RLM\quad- Empirical Coefficients of the Ridley Legacy Model}\\
		    \multicolumn{7}{l}{CMEE\quad- Empirical Coefficients of the Conductance Model for Extreme Events}\\
		    \multicolumn{7}{l}{RLM$^+$\quad- Ridley Legacy Model, with Auroral Oval Adjustments}\\
			\multicolumn{7}{l}{CMEE$^+$\quad- Conductance Model for Extreme Events, with Auroral Oval Adjustments}\\
			\hline
			\hline
		\end{tabular}
		\caption{
		Comparison of Heidke Skill Scores (HSS) for the space weather events listed in Table \ref{sims_table}(a) at the prescribed four $dB/dt$ thresholds (leftmost column) from \emph{Pulkkinen2013}. 
		(a) 
		The top-most table compares HSS for the conductance coefficients of RLM and CMEE; no auroral amelioration added to the model; 
		(b) The middle table compares results simulated using the CMEE using only the empirical conductance coefficients, against another version of the model that uses the CMEE coefficients along with the artificial oval adjustments;
		(c) 
		The bottom-most table compares the two empirical models with the auroral oval adjustments. 
		Here, green signifies improvement, while red signifies deterioration in prediction value.
		} 
		\label{hss_dbdt_table}
	\end{center}
\end{table}

\subsection{Performance Analysis of $\Delta B$ Estimation}
\label{dB_metric_section}

\begin{figure}[h!]
	\begin{center}
		\includegraphics[width=\textwidth]{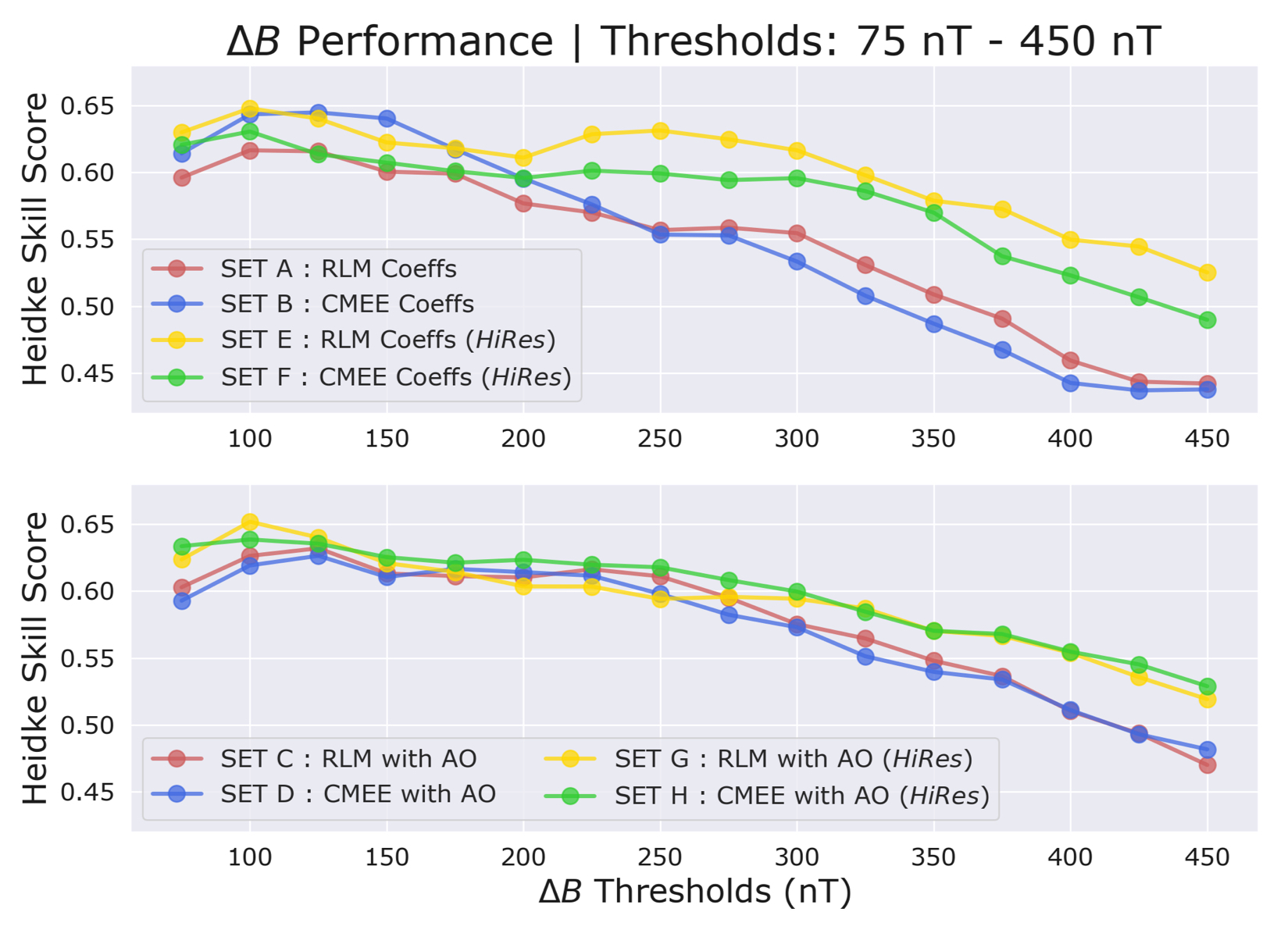}
	\end{center}
    \caption{HSS Performance metrics of all SWMF simulation variants at ascending $\Delta B$ predictions for all events from Table \ref{sims_table}(a). The format is similar to Figure \ref{hss_dbdt_plot}. Note the y-axis in (a) and (c) does not start at zero, and spans a smaller range than Figures \ref{hss_dbdt_plot}(a) and (c).}
    \label{hss_db_plot}
\end{figure}

Unlike the $dB/dt$ performance quantification using binary event analysis, the usage of the same procedure on $\Delta B$ values does not help address the science questions posed in Section \ref{intro}.
Figure \ref{hss_db_plot} describes 
variation in HSS for predicted $\Delta B$ from all model variants against observed values. 
In comparison to the $dB/dt$ predictions, the change in $\Delta B$ predictions were not nearly as drastic for better or worse. 
Note that the y-axis in Figures \ref{hss_db_plot}(a) and (b) are not the same as in Figures \ref{hss_dbdt_plot}(a) and (b); the HSS range spanned in the case of $\Delta B$ is much shorter than in the case of $dB/dt$. In part (a), 
the CMEE-driven predictions show deterioration in the HSS values compared to RLM. {\color{black}However, in comparison to the variation in HSS for $dB/dt$ by the expanded dataset, the variation observed is minimal.} 
The decrease in HSS values was similar, but lesser, in the \emph{Hi-Res} Set F results. 
In part (b), the variation in $\Delta B$ HSS values are negligible when oval adjustments were applied, for both model resolutions. 
In fact, 
some higher thresholds in part (b) showed no substantial change 
in the HSS values with the CMEE-driven simulations. 
When comparing 
parts (a) and (b) of Figure \ref{hss_db_plot}, the HSS values 
in part (b) are greater than their respective counterpart in part (a) of the figure for thresholds $\geq$ 200 $nT$. This indicates that while changing coefficients by increasing the dataset caused more variation in the HSS values of individual simulation sets, application of oval adjustments improves overall performance regardless of the coefficients used. 
For a more quantitative explanation of the $\Delta B$ performance, Table \ref{db_metric_table} presents values of all performance metrics calculated for all model variants at a high $\Delta B$ threshold of $400$ nT. 
The table is similarly structured to Table \ref{dbdt_metric_table} with the worst performance in each configuration coloured orange and the best performance coloured blue. 
When comparing the coefficient-driven simulations of RLM and CMEE,  
substantial variations are not observed in almost all skill scores with a maximum difference of $\sim 0.02$ for any given skill score and resolution. 
The same is seen with the simulations 
driven with oval adjustments, 
which also do not vary substantially. 
However, a significant jump is observed in the skill scores when comparing the impact of oval adjustments with oval adjusted simulations performing better than only coefficient-driven simulations. 
For both low and high res configurations, TS and F$_1$ skill scores improve when oval adjustments are applied. 
This is also seen in the accuracy measures like POD and MR whose values improve, with the POD jumping by a value of $\sim$0.1 indicating that the number of hits are increasing and number of misses decreasing. 
Similar to the $dB/dt$ metric analysis and in sharp contrast to the aforementioned performance metrics, the POFD and FAR values are best for simulations driven using non-oval adjustment applications.  
This is similar to the results in Section \ref{dBdt_metric_section}, where false alarms increase as we switch conductance models. 
Similar to Section \ref{dBdt_metric_section}, the trend seen in these performance metrics are not an isolated case for this specific threshold, but observed in all thresholds. The performance metrics for the other thresholds have been presented in the supp. material. 

The TSS and HSS do not show substantial differences 
as the conductance is modified, 
with the maximum difference between skill scores not being more than $\sim 0.05$. By comparison, the difference between the best and the worst HSS performance for the $dB/dt$ is $\sim 0.11$. The results also show that the best HSS and TSS for the \emph{Hi-Res} case are simulations driven by RLM coefficients, which is in direct contrast to the low res case where RLM coefficients consistently underperform for both TSS and HSS. 
%
This contrast is as a result of using the same time forecast window $t_f$ as the \emph{Pulkkinen2013} on $\Delta B$ predictions. The comparison window $t_f$ of 20 minutes, used in both this study and the \emph{Pulkkinen2013} study for $dB/dt$ predictions, is not long enough to observe severe variations in $\Delta B$ perturbations. As an example, the predicted $\Delta B$ hardly varies over more than two of the pre-determined thresholds, even during strong driving. 
In comparison, $dB/dt$ varies over multiple thresholds several times within a $t_f$. This shows that the metrics used in this study are not totally appropriate to study improvements in $\Delta B$ predictions. This could simply be done by increasing the comparison time window, or by using different error or bias metrics. As discussed earlier in Section \ref{qual_results} 
estimation of SSPB in Figure \ref{delb_fig} for specific magnetometer stations during Event 2 gives a quantitative understanding of the difference. 

\section{Analysis} \label{analysis}

The considerable increase in the frequency and magnitude of $dB/dt$ spikes at YKC with the application of the oval adjustments 
in Figure \ref{dbdt_fig}(b) is closely associated to the domain constraints in RIM. As described in Section \ref{rlm_coeffs}, while RIM's simulation domain 
spans the ionosphere pole-to-pole, 
the empirical auroral conductance module 
is limited with a spatial domain spanning the poles to MLat 60$^o$. 
This means that in its present configuration the auroral conductance module, be it RLM or CMEE, is bounded at MLat 60$^o$, with conductance values equatorward of this boundary dropping exponentially and the aurora being constrained poleward of the boundary. 
The impact of this boundary is clearly indicated in Figure \ref{delb_fig}(d), where auroral currents are the dominant source of ground $\Delta B$ in high latitude regions like YKC, 
but contribute negligibly at mid latitudinal regions like NEW. 
%

Since application of both the dataset expansion and oval adjustments result in increasing the conductance ceiling during strong driving, 
CMEE allows more magnetospheric currents to close more dynamically throughout the ionosphere 
at any given time. In addition, the oval adjustments enhance conductance in regions of high upward FACs thereby changing the pattern of the auroral conductance and reducing the conductance as a function of distance from the empirically constructed oval. The combined effect of these modifications would result in the auroral horizontal currents in RIM's domain being estimated with increased accuracy. 
This, in turn, leads to a more accurate estimation of the $\Delta B$ perturbation and subsequently $dB/dt$ 
, which are both 
calculated from the Biot-Savart integral of these current systems (e.g. \citealp{Yu2010, Welling2019}). 
The conductance modifications due to the two elements (dataset expansion and oval adjustment) lead to noisier results in $dB/dt$, which 
leads to increased 
spikes. These spikes, when correct, increase the number of hits and when incorrect, increase the number of false alarms. 
The emergence of $dB/dt$ spikes in the modeled data during 
the oval expansion phase in the bottom subplot of Figure \ref{dbdt_fig}(b) demarcates why false alarms increase when the oval adjustment factor is used. 
In addition to the boundary constraints, false alarms are also caused by sudden shifting of the empirically-estimated auroral oval. 
These shifts 
are caused as a result of the sensitive dependence of the oval adjustments to changes in FAC patterns. Sharp changes in the FAC occuring over time scales in the same order of the coupling time cadence cause the empirical estimation of the oval to change rapidly. 
This brisk movement of the placement of the oval adjustment results in the loci movement of $dB/dt$ spikes, 
causing unexpected hits and/or false alarms. In all, the aforementioned problems place the auroral oval in the wrong spot which lead to $dB/dt$ spikes, perhaps even at the right time, but wrong location hence increasing the false alarms.

While an increment in the number of false alarms is a significant drawback, the advantages of using the improved conductance model in the SWMF far outweigh this issue. Firstly, the expansion of the dataset in CMEE allows for an increased limit cap on the magnitude of the conductance which results in generating a more realistic cross polar cap potential to be fed back as input to the GM and IM modules. This is essential when conducting numerical experiments investigating the magnetosphere-ionosphere coupling. Secondly, the changes in the conductance pattern in CMEE, as a result of the use of nonlinear regression, 
physically alters the nightside and dayside auroral conductance pattern when compared to RLM. 
Using global modeling, this numerical experiment has not only been able to address the question of expanded dataset raised by \emph{Welling2017}, but is also able to discern the impact of ionospheric conductance on space weather forecasting. Finally, both the magnitude and pattern of ionospheric conductance proves to be an important quantity in affecting a global model's $dB/dt$ predictive skill. Given that the $dB/dt$ is an important quantity used in the science community and the industry to predict space weather on the ground, accuracy in the ionospheric conductance is important in our global models. Through this work, the authors present an advanced and more accurate auroral conductance model to address this challenge.

\begin{table}
    \centering
    \begin{tabular}{c | c c c c|c c c c}
        \multirow{2}{*}{Metric}& \multicolumn{4}{c}{\emph{SWPC} Configuration} & \multicolumn{4}{c}{\emph{Hi-Res SWPC} Configuration}\\
        & RLM 
        & CMEE 
        & RLM$^+$ 
        & CMEE$^+$ 
        & RLM 
        & CMEE 
        & RLM$^+$ 
        & CMEE$^+$ 
        \\
        \hline
        POD     &\textcolor{orange}{$0.2216$}       &$0.2490$       &$0.2668$       &\textcolor{blue}{$0.3557$}       &\textcolor{orange}{$0.2791$}       &$0.3406$       &$0.4309$       &\textcolor{blue}{$0.5554$} \\
        POFD     &\textcolor{blue}{$0.0169$}       &$0.0194$       &$0.0253$       &\textcolor{orange}{$0.0319$}       &\textcolor{blue}{$0.0262$}       &$0.0378$       &$0.0566$       &\textcolor{orange}{$0.0784$} \\
        FAR     &\textcolor{blue}{$0.3306$}       &$0.3358$       &\textcolor{orange}{$0.3810$}       &$0.3674$       &\textcolor{blue}{$0.3780$}       &$0.4182$       &$0.4597$       &\textcolor{orange}{$0.4775$} \\
        MR      &\textcolor{orange}{$0.1089$}       &$0.1057$       &$0.1041$       &\textcolor{blue}{$0.0932$}       &\textcolor{orange}{$0.1026$}       &$0.0957$       &$0.0852$       &\textcolor{blue}{$0.0693$} \\
        \hline
        TS      &\textcolor{orange}{$0.1998$}       &$0.2211$       &$0.2291$       &\textcolor{blue}{$0.2948$}       &\textcolor{orange}{$0.2386$}       &$0.2736$       &$0.3153$       &\textcolor{blue}{$0.3684$} \\
        F1      &\textcolor{orange}{$0.3330$}       &$0.3622$       &$0.3728$       &\textcolor{blue}{$0.4553$}       &\textcolor{orange}{$0.3853$}       &$0.4297$       &$0.4795$       &\textcolor{blue}{$0.5385$} \\
        TSS     &\textcolor{blue}{$0.5605$}       &$0.5585$       &\textcolor{orange}{$0.5150$}       &$0.5394$       &\textcolor{blue}{$0.5194$}       &$0.4861$       &$0.4551$       &\textcolor{orange}{$0.4532$} \\
        HSS     &\textcolor{orange}{$0.2855$}       &$0.3120$       &$0.3179$       &\textcolor{blue}{$0.3973$}       &\textcolor{orange}{$0.3297$}       &$0.3672$       &$0.4094$       &\textcolor{blue}{$0.4647$} \\
        \hline
        \end{tabular}
    \caption{Performance metrics table for predicted $dB/dt$ at the $1.5$ $nT/s$ threshold. Listed are all performance metrics defined in Table \ref{metric_tab} (Leftmost column) measured for SWMF simulations conducted using RLM Coefficients (denoted by '\textit{RLM}'), CMEE Coefficients (denoted by '\textit{CMEE}'), RLM with oval adjustment (denoted by '\textit{RLM$^+$}') and CMEE with oval adjustment (denoted by '\textit{CMEE}$^+$') simulated using both the \emph{SWPC} and \emph{Hi-Res SWPC} configurations. The orange values show the least desirable metric results, while the blue values signify the best results for this threshold.} \label{dbdt_metric_table}
\end{table}
\begin{table}
    \centering
    \begin{tabular}{c | c c c c|c c c c}
        \multirow{2}{*}{Metric}& \multicolumn{4}{c}{\emph{SWPC} Configuration} & \multicolumn{4}{c}{\emph{Hi-Res SWPC} Configuration}\\
        & RLM 
        & CMEE 
        & RLM$^+$ 
        & CMEE$^+$ 
        & RLM 
        & CMEE 
        & RLM$^+$ 
        & CMEE$^+$ 
        \\
        \hline
        POD     &$0.4602$       &\textcolor{orange}{$0.4385$}       &$0.5123$       &\textcolor{blue}{$0.5224$}       &$0.5687$       &\textcolor{orange}{$0.5485$}       &$0.6440$       &\textcolor{blue}{$0.6671$} \\
        POFD     &$0.0575$       &\textcolor{blue}{$0.0523$}       &$0.0616$       &\textcolor{orange}{$0.0658$}       &\textcolor{blue}{$0.0865$}       &$0.0901$       &$0.1393$       &\textcolor{orange}{$0.1429$} \\
        FAR     &$0.2587$       &\textcolor{blue}{$0.2500$}       &$0.2516$       &\textcolor{orange}{$0.2602$}       &\textcolor{blue}{$0.2982$}       &$0.3146$       &\textcolor{orange}{$0.3768$}       &$0.3745$ \\
        MR      &$0.1701$       &\textcolor{orange}{$0.1749$}       &$0.1568$       &\textcolor{blue}{$0.1546$}       &$0.1445$       &\textcolor{orange}{$0.1508$}       &$0.1289$       &\textcolor{blue}{$0.1220$} \\
        \hline
        TS      &$0.3965$       &\textcolor{orange}{$0.3826$}       &$0.4370$       &\textcolor{blue}{$0.4413$}       &$0.4580$       &\textcolor{orange}{$0.4382$}       &$0.4635$       &\textcolor{blue}{$0.4767$} \\
        F1      &$0.5679$       &\textcolor{orange}{$0.5534$}       &$0.6082$       &\textcolor{blue}{$0.6124$}       &$0.6283$       &\textcolor{orange}{$0.6093$}       &$0.6335$       &\textcolor{blue}{$0.6457$} \\
        TSS     &\textcolor{orange}{$0.5712$}       &$0.5751$       &\textcolor{blue}{$0.5916$}       &$0.5851$       &\textcolor{blue}{$0.5573$}       &$0.5346$       &\textcolor{orange}{$0.4943$}       &$0.5035$ \\
        HSS     &$0.4585$       &\textcolor{orange}{$0.4456$}       &$0.5015$       &\textcolor{blue}{$0.5042$}       &\textcolor{blue}{$0.5135$}       &\textcolor{orange}{$0.4898$}       &$0.4994$       &$0.5132$ \\
        \hline
    \end{tabular}
    \caption{Performance metrics table for predicted $\Delta B$ at the $400$ $nT$ threshold. Listed are all performance metrics defined in Table \ref{metric_tab} (Leftmost column) measured for SWMF simulations conducted using the same variants as in Table \ref{dbdt_metric_table}. The orange values show the least desirable metric results, while the blue values signify the best results for this threshold.} \label{db_metric_table}
\end{table}

\section{Conclusion}

In this work, the development of an advanced auroral conductance model, CMEE has been presented. CMEE has been designed using nonlinear regression to span minute-resolution data generated from AMIE for the whole year of 2003 spanning extreme events. It has additional capability to add physics-driven empirical adjustments to improve the auroral conductance 
to ensure a larger range on conductance values to better predict the conductance for a broad range of activity. In this study, this model has been used in the SWMF 
to investigate the impact of auroral conductance on space weather prediction. 
Simulated results were compared against observed global quantities like polar cap potential, field aligned current intensity and ground-based magnetic perturbation. Additionally, a quantitative investigation was conducted using a binary event analysis similar to the \emph{Pulkkinen2013} study and skill scores for $dB/dt$ and $\Delta B$ predictions were computed.

The investigation showed that application of the increased dataset coupled with oval adjustments led to substantial changes in almost all space weather quantities. CMEE allows the auroral conductance to have an increased range of values, attaining a higher ceiling during extreme driving as compared to RLM. 
Since FACs are largely driven by upstream conditions, they were not drastically impacted by changes in the conductance model. However, since the conductance value increased and FACs varied minimally, the CPCP values were lowered with the usage of CMEE and the oval adjustments. Since, auroral horizontal currents directly impact the ground magnetic perturbation $\Delta B$ and its temporal variant $dB/dt$, the driving of both these quantities were appreciably altered by the application of both the expanded dataset and oval adjustments. While usage of the expanded dataset resulted in a general increase of the modeled $dB/dt$ magnitude, oval adjustments increased the frequency of $dB/dt$ spikes. Neither of these properties were able to improve the modeling of the auroral oval expansion. This resulted in the formation of different regimes in the latitudinal contribution to the $\Delta B$ and $dB/dt$ distributions, with negligble contribution of auroral currents in low or mid latitude magnetometer stations in the modeled output during extreme driving. 

The results of the binary event analysis conducted on the 
simulation variants indicated that usage of CMEE with oval adjustments yields best performance, with drastic improvements in the HSS metric  at higher activity thresholds. In addition, most performance metrics exhibited favourable changes when applying the CMEE coefficeints and/or oval adjustments, 
indicating an increase in the number of identified hits and true negatives and a decrease in misses. However, the performance metrics also indicated that the number of false alarms increased with the application of the oval adjustment. This was caused predominantly because of the brisk movement of the empirically-estimated oval, and  the latitudinal constraint on the auroral conductance which inhibits the oval from expanding beyond MLat 60$^\circ$, thereby pushing the auroral currents poleward. 
While this process increases the number of hits, favourably affecting most performance metrics, it also hurts metrics like TSS due to increased number of false alarms. The binary event analysis of $\Delta B$ predictions do not yield definitive results, exhibiting minimal impact on skill scores. This is most likely because the time forecast window of 20 minutes, chosen to study $dB/dt$ forecasts in the original \emph{Pulkkinen2013} study, is limited for the $\Delta B$ to exhibit significant change in value so as to jump multiple number of thresholds and therefore produce any meaningful changes in the performance metrics.
Outstanding shortcomings of the present analysis such as those mentioned above and additional analysis like estimation of bias and error metrics for various thresholds 
are steps that we are presently pursuing. {{In addition, a key drawback of the present method is that the method of estimating the conductance using AMIE data from times of extreme driving is inconsistent, since the auroral conductance in AMIE is itself derived using an empirical relationship} (\citealp{Ahn1998}){.}} Because validation is a process, continued data-model comparisons will be performed in future studies. {{Further comparisons of the conductance estimates, field aligned current and potential patterns against measurements by AMIE, SuperDARN and DMSP crossings will be presented.}}

The issues causing the misidentification of $dB/dt$ spikes 
requires a physical solution with numerical modifications to allow the aurora to expand to mid or low latitudes during extreme events. 
While this could be done with data, an easier and more novel solution would be to drive precipitation from the magnetospheric domains. This could be done by coupling physics-based precipitative inputs from GM and IM modules to estimate electron and ion precipitation in the aurora. This is similar to what has been done in studies like \cite{Raeder2001} and \cite{Wiltberger2009}. Such an approach allows for a novel approach to isolate and understand the impact of individual sources of auroral conductance. At the same time, the precipitation pattern of the aurora allows observational data from extreme events to feature prominently in perceiving the accuracy of precipitative fluxes at different MLTs and magnetic latitudes. The development of such a model is presently being undertaken by the authors to address the aforementioned {{issues of dataset inconsistencies and oval expansion}} \citep{Mukhopadhyay2018AGU, Mukhopadhyay2019AGU}.

In conclusion, the usage of CMEE designed using an increased dataset coupled with the application of oval adjustment parameters lead to substantial changes in our $dB/dt$ predictions. 
With the crucial impact that the auroral conductance imparts on global quantities, CMEE would serve as a competent replacement to RLM's coefficient map. The usage of the oval adjustments in the SWMF's auroral conductance estimation is unique and compelling in driving future developments of auroral conductance models to acheive accuracy in the conductance pattern, in addition to the magnitude. Additionally, as evidenced by the skill score analysis, the new model leads to significant improvement in predictive skill of our space weather model.

\acknowledgments
Support for this work has been provided by NASA Grants NNX17AB87G, 80NSSC18K1120, and 80NSSC17K0015, and NSF Grant 1663770. We would like to acknowledge high-performance computing support from Pleaides (
allocation 1815) provided by NASA's High-End Computing Capability Programme, and Cheyenne (
allocation UUSL0016) provided by NCAR's Computational and Information Systems Laboratory, sponsored by the National Science Foundation. Model result data, input files and observation data are available via \href{https://doi.org/10.7302/nwxp-g551}{https://doi.org/10.7302/nwxp-g551}. The Space Weather Modeling Framework is maintained by the University of Michigan Center for Space Environment Modeling and can be obtained at \href{http://csem.engin.umich.edu/tools/swmf}{http://csem.engin.umich.edu/tools/swmf}. AMIE Results used in this study are maintained at the University of Michigan's Virtual Model Repository (VMR; \href{http://vmr.engin.umich.edu/}{http://vmr.engin.umich.edu/}). {{The authors thank NASA Community Coordinated Modeling Center (CCMC) Staff for providing the magnetometer measurements.}}

The authors would like to thank Dr. Meghan Burleigh for reading a draft manuscript. We thank Dr. Shasha Zou, Dr. Robert Robinson, Dr. Steven Morley and Dr. Gabor Toth for sharing their expertise in the course of this study. A.M. would like to thank Dr. Dogacan su Ozturk, Dr. Zhenguang Huang, Dr. Natalia Ganjushkina, 
Ms. Abigail Azari, Mr. Alexander Shane, 
Mr. Brian Swiger and Mr. Christopher Bert for sharing their expertise during the development of modeling, curve-fitting and validation tools used in this study.


%
%

\bibliography{cmee_paper.bib}

%
%
%
%
%

\end{document}


%
%


\title{Supporting Information for "Conductance Model for Extreme Events : Impact of Auroral Conductance on Space Weather Forecasts"}
%
%

%
%



\authors{Agnit Mukhopadhyay\affil{1}, Daniel T. Welling\affil{2}, Michael W. Liemohn\affil{1}, Aaron J. Ridley\affil{1}, Shibaji Chakraborty\affil{3}, and Brian J. Anderson\affil{4}}


\affiliation{1}{Climate and Space Sciences and Engineering Department, University of Michigan, Ann Arbor, MI, USA}
\affiliation{2}{Department of Physics, University of Texas at Arlington, Arlington, TX, USA}
\affiliation{3}{Department of Electrical and Computer Engineering, Virginia Polytechnic Institute and State University, Blacksburg, VA, USA}
\affiliation{4}{Applied Physics Laboratory, Johns Hopkins University, Baltimore, MD, USA}

%
%

%

\begin{article}

%
%

\noindent\textbf{Contents of this file}
\begin{enumerate}
\item Tables S1 to S15
\end{enumerate}

\noindent\textbf{Introduction}
This supporting information provides performance metrics calculated for multiple $dB/dt$ and $\Delta B$ thresholds using the Conductance Model for Extreme Events. The metrics used has been listed in Table 2 of the main article. The format of these tables are similar to Tables 4 and 5 of the main article; for more details about those tables, please refer to Sections 3.2 and 3.3 of the main paper. For convenience, the tables have been coloured differently: In the tables, \textit{\textbf{italicized-bolded} text} is used to denote best performance while \textit{\underline{italicized-underlined} text} is used to denote worst. Usage of the auroral oval and CMEE amounts to an increase in False Negatives (F) in both $\Delta B$ and $dB/dt$ predictions, while improving the rest of the quantities (H, M, N). Due to this reason, the FAR values are higher for oval runs, which results in less predictive score using the TSS metric.
The new model (without the oval) has more misses (M) than the older model (without the oval), when predicting $\Delta B$. For $dB/dt$ predictions, the amount of skill lost during quieter activity, when simulating using CMEE, is more than regained with massive improvements for extreme driving, as is seen by Tables S3 to S7.


\noindent\textbf{Table S1. Performance metrics for predicted $dB/dt$ at Threshold = \textbf{0.1 nT/s}.}

\centering
\begin{array}{c | c c c c|c c c c}
        Metric & SET A & SET B & SET C & SET D &
        SET E & SET F & SET G & SET H\\
        \hline
        POD     &\textit{\underline{0.7776}}       &$0.7801$       &\textbf{\textit{0.8436}}       &$0.8114$       &\textit{\underline{0.8342}}       &$0.8555$       &\textit{\textbf{0.8976}}       &$0.8973$ \\
        POFD     &$0.0631$       &\textbf{\textit{0.0615}}       &\textit{\underline{0.0903}}       &$0.0744$       &$0.0938$       &\textit{\textbf{0.0892}}       &$0.1687$       &\textit{\underline{0.1697}} \\
        FAR    &$0.0431$       &\textbf{\textit{0.0419}}       &\underline{\textit{0.0560}}       &$0.0484$       &$0.0587$       &\textit{\textbf{0.0547}}       &$0.0944$       &\textit{\underline{0.0950}} \\
        MR      &\textit{\underline{0.2997}}       &$0.2970$       &\textit{\textbf{0.2367}}       &$0.2686$       &\textit{\underline{0.2481}}       &$0.2224$       &\textit{\textbf{0.1817}}       &$0.1823$ \\
\end{array}

\vspace{2em}

\begin{flushleft}
\noindent\textbf{Table S2. Performance metrics for predicted $dB/dt$ at Threshold = \textbf{0.3 nT/s}.}
\end{flushleft}

\begin{array}{c | c c c c|c c c c}
        Metric & SET A & SET B & SET C & SET D &
        SET E & SET F & SET G & SET H\\
        \hline
        POD     &$0.5814$       &\textit{\underline{0.5628}}       &\textit{\textbf{0.6782}}       &$0.6638$       &\textit{\underline{0.6444}}       &$0.6829$       &$0.7970$       &\textit{\textbf{0.7979}} \\
        POFD     &$0.0416$       &\textit{\textbf{0.0345}}       &\textit{\underline{0.0597}}       &$0.0477$       &\textit{\textbf{0.0426}}       &$0.0613$       &$0.1038$       &\textit{\underline{0.1164}} \\
        FAR     &$0.0858$       &\textit{\textbf{0.0744}}       &\textit{\underline{0.1034}}       &$0.0861$       &\textit{\textbf{0.0797}}       &$0.1053$       &$0.1459$       &\textit{\underline{0.1606}} \\
        MR      &$0.2499$       &\textit{\underline{0.2567}}       &\textit{\textbf{0.2070}}       &$0.2121$       &\textit{\underline{0.2207}}       &$0.2049$       &\textit{\textbf{0.1473}}       &$0.1485$ \\
\end{array}
\vspace{2em}


\begin{flushleft}
\noindent\textbf{Table S3. Performance metrics for predicted $dB/dt$ at Threshold = \textbf{0.5 nT/s}.}
\end{flushleft}

\begin{array}{c | c c c c|c c c c}
        Metric & SET A & SET B & SET C & SET D &
        SET E & SET F & SET G & SET H\\
        \hline
        POD     &\textit{\underline{0.4812}}       &$0.4840$       &$0.5636$       &\textit{\textbf{0.5922}}      &\textit{\underline{0.5496}}       &$0.5989$       &$0.7244$       &\textit{\textbf{0.7451}} \\
        POFD     &$0.0331$       &\textit{\textbf{0.0302}}       &\textit{\underline{0.0497}}       &$0.0435$       &\textit{\textbf{0.0378}}       &$0.0543$       &$0.0926$       &\textit{\underline{0.1076}} \\
        FAR     &$0.1244$       &\textit{\textbf{0.1138}}       &\textit{\underline{0.1539}}       &$0.1315$       &\textit{\textbf{0.1241}}       &$0.1576$       &$0.2087$       &\textit{\underline{0.2294}} \\
        MR      &\textit{\underline{0.2065}}       &$0.2051$       &$0.1821$       &\textit{\textbf{0.1713}}       &\textit{\underline{0.1850}}       &$0.1706$       &$0.1284$       &\textit{\textbf{0.1217}} \\
\end{array}
\vspace{2em}


\begin{flushleft}
\noindent\textbf{Table S4. Performance metrics for predicted $dB/dt$ at Threshold = \textbf{0.7 nT/s}.}
\end{flushleft}

\begin{array}{c | c c c c|c c c c}
        Metric & SET A & SET B & SET C & SET D &
        SET E & SET F & SET G & SET H\\
        \hline
        POD     &\textit{\underline{0.3949}}       &$0.4278$       &$0.4661$       &\textit{\textbf{0.5305}}       &\textit{\underline{0.4812}}       &$0.5441$       &$0.6537$       &\textit{\textbf{0.7057}} \\
        POFD     &\textit{\textbf{0.0255}}       &$0.0260$       &$0.0365$       &\textit{\underline{0.0387}}       &\textit{\textbf{0.0285}}       &$0.0442$       &$0.0801$       &\textit{\underline{0.0936}} \\
        FAR     &$0.1502$       &\textit{\textbf{0.1427}}       &\textit{\underline{0.1765}}       &$0.1667$       &\textit{\textbf{0.1395}}       &$0.1821$       &$0.2516$       &\textit{\underline{0.2667}} \\
        MR      &\textit{\underline{0.1847}}       &$0.1765$       &$0.1681$       &\textit{\textbf{0.1512}}       &\textit{\underline{0.1630}}       &$0.1482$       &$0.1208$       &\textit{\textbf{0.1059}} \\
\end{array}
\vspace{2em}


\begin{flushleft}
\noindent\textbf{Table S5. Performance metrics for predicted $dB/dt$ at Threshold = \textbf{1.1 nT/s}.}
\end{flushleft}

\begin{array}{c | c c c c|c c c c}
        Metric & SET A & SET B & SET C & SET D &
        SET E & SET F & SET G & SET H\\
        \hline
        POD     &\textit{\underline{0.2920}}       &$0.3310$       &$0.3520$       &\textit{\textbf{0.4380}}       &\textit{\underline{0.3770}}       &$0.4440$       &$0.5300$       &\textit{\textbf{0.6230}} \\
        POFD     &\textit{\textbf{0.0222}}       &$0.0235$       &$0.0284$       &\textit{\underline{0.0372}}       &\textit{\textbf{0.0273}}       &$0.0434$       &$0.0687$       &\textit{\underline{0.0914}} \\
        FAR     &$0.2532$       &\textit{\textbf{0.2408}}       &$0.2651$       &\textit{\underline{0.2748}}       &\textit{\textbf{0.2445}}       &$0.3041$       &$0.3668$       &\textit{\underline{0.3957}} \\
        MR      &\textit{\underline{0.1395}}       &$0.1330$       &$0.1299$       &\textit{\textbf{0.1156}}       &\textit{\underline{0.1254}}       &$0.1152$       &$0.1015$       &\textit{\textbf{0.0850}} \\
\end{array}
\vspace{2em}


\begin{flushleft}
\noindent\textbf{Table S6. Performance metrics for predicted $dB/dt$ at Threshold = \textbf{1.5 nT/s}.}
\end{flushleft}

\begin{array}{c | c c c c|c c c c}
        Metric & SET A & SET B & SET C & SET D &
        SET E & SET F & SET G & SET H\\
        \hline
        POD     &\textit{\underline{0.2216}}       &$0.2490$       &$0.2668$       &\textit{\textbf{0.3557}}       &\textit{\underline{0.2791}}       &$0.3406$       &$0.4309$       &\textit{\textbf{0.5554}} \\
        POFD     &\textit{\textbf{0.0169}}       &$0.0194$       &$0.0253$       &\textit{\underline{0.0319}}       &\textit{\textbf{0.0262}}       &$0.0378$       &$0.0566$       &\textit{\underline{0.0784}} \\
        FAR     &\textit{\textbf{0.3306}}       &$0.3358$       &\textit{\underline{0.3810}}       &$0.3674$       &\textit{\textbf{0.3780}}       &$0.4182$       &$0.4597$       &\textit{\underline{0.4775}} \\
        MR      &\textit{\underline{0.1089}}       &$0.1057$       &$0.1041$       &\textit{\textbf{0.0932}}       &\textit{\underline{0.1026}}       &$0.0957$       &$0.0852$       &\textit{\textbf{0.0693}} \\
\end{array}
\vspace{2em}


\begin{flushleft}
\noindent\textbf{Table S7. Performance metrics for predicted $dB/dt$ at Threshold = \textbf{1.7 nT/s}.}
\end{flushleft}

\begin{array}{c | c c c c|c c c c}
        Metric & SET A & SET B & SET C & SET D &
        SET E & SET F & SET G & SET H\\
        \hline
        POD     &\textit{\underline{0.1975}}       &$0.2070$       &$0.2389$       &\textit{\textbf{0.3264}}       &\textit{\underline{0.2452}}       &$0.3153$       &$0.3838$       &\textit{\textbf{0.5207}} \\
        POFD     &$0.0161$       &\textit{\textbf{0.0155}}       &$0.0238$       &\textit{\underline{0.0287}}       &\textit{\textbf{0.0246}}       &$0.0356$       &$0.0529$       &\textit{\underline{0.0754}} \\
        FAR     &$0.3861$       &\textit{\textbf{0.3659}}       &\textit{\underline{0.4340}}       &$0.4041$       &\textit{\textbf{0.4359}}       &$0.4649$       &$0.5151$       &\textit{\underline{0.5275}} \\
        MR      &\textit{\underline{0.0957}}       &$0.0947$       &$0.0919$       &\textit{\textbf{0.0826}}       &\textit{\underline{0.0913}}       &$0.0844$       &$0.0779$       &\textit{\textbf{0.0630}} \\
\end{array}
\vspace{2em}




\begin{flushleft}
\noindent\textbf{Table S8. Performance metrics for predicted $\Delta B$ at Threshold = \textbf{100 nT}.}
\end{flushleft}

\begin{array}{c | c c c c|c c c c}
        Metric & SET A & SET B & SET C & SET D &
        SET E & SET F & SET G & SET H\\
        \hline
        POD     &\textit{\underline{0.7293}}       &$0.7918$       &\textit{\textbf{0.8099}}       &$0.7879$       &$0.8603$       &\textit{\underline{0.8373}}       &\textit{\textbf{0.8882}}       &$0.8740$ \\
        POF     &\textit{\textbf{0.1882}}       &$0.2296$       &\textit{\underline{0.3074}}       &$0.2836$       &$0.3388$       &\textit{\textbf{0.3199}}       &\textit{\underline{0.4442}}       &$0.4003$ \\
        FAR     &\textit{\textbf{0.1013}}       &$0.1124$       &\textit{\underline{0.1422}}       &$0.1358$       &$0.1467$       &\textit{\textbf{0.1430}}       &\textit{\underline{0.1792}}       &$0.1667$ \\
        MR      &\textit{\underline{0.4330}}       &\textit{\textbf{0.3823}}       &$0.3860$       &$0.4040$       &$0.3261$       &\textit{\underline{0.3540}}       &\textit{\textbf{0.3153}}       &$0.3249$ \\
\end{array}
\vspace{2em}


\begin{flushleft}
\noindent\textbf{Table S9. Performance metrics for predicted $\Delta B$ at Threshold = \textbf{150 nT}.}
\end{flushleft}

\begin{array}{c | c c c c|c c c c}
        Metric & SET A & SET B & SET C & SET D &
        SET E & SET F & SET G & SET H\\
        \hline
        POD     &\textit{\underline{0.6803}}       &$0.7440$       &\textit{\textbf{0.7534}}       &$0.7460$       &$0.8184$       &\textit{\underline{0.7842}}       &\textit{\textbf{0.8599}}       &$0.8432$ \\
        POF     &\textit{\textbf{0.1186}}       &$0.1460$       &\textit{\underline{0.2487}}       &$0.2186$       &$0.2743$       &\textit{\textbf{0.2469}}       &\textit{\underline{0.3823}}       &$0.3354$ \\
        FAR     &\textit{\textbf{0.1166}}       &$0.1294$       &\textit{\underline{0.2000}}       &$0.1816$       &$0.2025$       &\textit{\textbf{0.1925}}       &\textit{\underline{0.2519}}       &$0.2315$ \\
        MR      &\textit{\underline{0.3238}}       &\textit{\textbf{0.2836}}       &$0.3024$       &$0.3003$       &$0.2484$       &\textit{\underline{0.2745}}       &\textit{\textbf{0.2304}}       &$0.2376$ \\
\end{array}
\vspace{2em}


\begin{flushleft}
\noindent\textbf{Table S10. Performance metrics for predicted $\Delta B$ at Threshold = \textbf{200 nT}.}
\end{flushleft}

\begin{array}{c | c c c c|c c c c}
        Metric & SET A & SET B & SET C & SET D &
        SET E & SET F & SET G & SET H\\
        \hline
        POD     &\textit{\underline{0.6331}}       &$0.6846$       &\textit{\textbf{0.7241}}       &$0.7128$       &$0.7812$       &\textit{\underline{0.7466}}       &\textit{\textbf{0.8142}}       &$0.8045$ \\
        POF     &\textit{\textbf{0.0711}}       &$0.1117$       &\textit{\underline{0.1697}}       &$0.1523$       &$0.2059$       &\textit{\textbf{0.1929}}       &\textit{\underline{0.3183}}       &$0.2683$ \\
        FAR     &\textit{\textbf{0.1107}}       &$0.1532$       &\textit{\underline{0.2063}}       &$0.1916$       &$0.2263$       &\textit{\textbf{0.2228}}       &\textit{\underline{0.3025}}       &$0.2701$ \\
        MR      &\textit{\underline{0.2625}}       &$0.2424$       &\textit{\textbf{0.2305}}       &$0.2339$       &$0.1990$       &\textit{\underline{0.2206}}       &$0.1973$       &\textit{\textbf{0.1941}} \\
\end{array}
\vspace{2em}


\begin{flushleft}
\noindent\textbf{Table S11. Performance metrics for predicted $\Delta B$ at Threshold = \textbf{250 nT}.}
\end{flushleft}

\begin{array}{c | c c c c|c c c c}
        Metric & SET A & SET B & SET C & SET D &
        SET E & SET F & SET G & SET H\\
        \hline
        POD     &\textit{\underline{0.5774}}       &$0.6071$       &\textit{\textbf{0.6998}}       &$0.6692$       &$0.7686$       &\textit{\underline{0.7180}}       &$0.7830$       &\textit{\textbf{0.7887}} \\
        POF     &\textit{\textbf{0.0527}}       &$0.0799$       &\textit{\underline{0.1231}}       &$0.1129$       &\textit{\textbf{0.1504}}       &\textit{\textbf{0.1504}}       &\textit{\underline{0.2621}}       &$0.2265$ \\
        FAR     &\textit{\textbf{0.1208}}       &$0.1656$       &\textit{\underline{0.2095}}       &$0.2027$       &\textit{\textbf{0.2277}}       &$0.2399$       &\textit{\underline{0.3352}}       &$0.3020$ \\
        MR      &\textit{\underline{0.2284}}       &$0.2208$       &\textit{\textbf{0.1851}}       &$0.1984$       &\textit{\textbf{0.1531}}       &\textit{\underline{0.1805}}       &$0.1633$       &$0.1535$ \\
\end{array}
\vspace{2em}


\begin{flushleft}
\noindent\textbf{Table S12. Performance metrics for predicted $\Delta B$ at Threshold = \textbf{300 nT}.}
\end{flushleft}

\begin{array}{c | c c c c|c c c c}
        Metric & SET A & SET B & SET C & SET D &
        SET E & SET F & SET G & SET H\\
        \hline
        POD     &$0.5553$       &\textit{\underline{0.5422}}       &\textit{\textbf{0.6254}}       &$0.6145$       &$0.7021$       &\textit{\underline{0.6703}}       &$0.7393$       &\textit{\textbf{0.7426}} \\
        POF     &\textit{\textbf{0.0456}}       &$0.0562$       &\textit{\underline{0.0960}}       &$0.0866$       &\textit{\textbf{0.1194}}       &$0.1229$       &\textit{\underline{0.2054}}       &$0.1949$ \\
        FAR     &\textit{\textbf{0.1333}}       &$0.1624$       &\textit{\underline{0.2231}}       &$0.2087$       &\textit{\textbf{0.2414}}       &$0.2555$       &\textit{\underline{0.3421}}       &$0.3294$ \\
        MR      &$0.1993$       &\textit{\underline{0.2058}}       &\textit{\textbf{0.1812}}       &$0.1840$       &$0.1531$       &\textit{\underline{0.1672}}       &$0.1491$       &\textit{\textbf{0.1459}} \\
\end{array}
\vspace{2em}


\begin{flushleft}
\noindent\textbf{Table S13. Performance metrics for predicted $\Delta B$ at Threshold = \textbf{350 nT}.}
\end{flushleft}

\begin{array}{c | c c c c|c c c c}
        Metric & SET A & SET B & SET C & SET D &
        SET E & SET F & SET G & SET H\\
        \hline
        POD     &$0.5076$       &\textit{\underline{0.4810}}       &\textit{\textbf{0.5711}}       &$0.5698$       &$0.6345$       &\textit{\underline{0.6193}}       &$0.6904$       &\textit{\textbf{0.6967}} \\
        POF     &$0.0529$       &\textit{\textbf{0.0507}}       &\textit{\underline{0.0752}}       &$0.0725$       &\textit{\textbf{0.1041}}       &$0.1052$       &\textit{\underline{0.1750}}       &$0.1696$ \\
        FAR     &\textit{\textbf{0.1952}}       &$0.1970$       &\textit{\underline{0.2347}}       &$0.2285$       &\textit{\textbf{0.2764}}       &$0.2834$       &\textit{\underline{0.3711}}       &$0.3616$ \\
        MR      &$0.1826$       &\textit{\underline{0.1902}}       &\textit{\textbf{0.1662}}       &\textit{\textbf{0.1662}}       &$0.1491$       &\textit{\underline{0.1546}}       &$0.1389$       &\textit{\textbf{0.1356}} \\
\end{array}
\vspace{2em}


\begin{flushleft}
\noindent\textbf{Table S14. Performance metrics for predicted $\Delta B$ at Threshold = \textbf{400 nT}.}
\end{flushleft}

\begin{array}{c | c c c c|c c c c}
        Metric & SET A & SET B & SET C & SET D &
        SET E & SET F & SET G & SET H\\
        \hline
        POD     &$0.4602$       &\textit{\underline{0.4385}}       &$0.5123$       &\textit{\textbf{0.5224}}       &$0.5687$       &\textit{\underline{0.5485}}       &$0.6440$       &\textit{\textbf{0.6671}} \\
        POF     &$0.0575$       &\textit{\textbf{0.0523}}       &$0.0616$       &\textit{\underline{0.0658}}       &\textit{\textbf{0.0865}}       &$0.0901$       &$0.1393$       &\textit{\underline{0.1429}} \\
        FAR     &$0.2587$       &\textit{\textbf{0.2500}}       &$0.2516$       &\textit{\underline{0.2602}}       &\textit{\textbf{0.2982}}       &$0.3146$       &\textit{\underline{0.3768}}       &$0.3745$ \\
        MR      &$0.1701$       &\textit{\underline{0.1749}}       &$0.1568$       &\textit{\textbf{0.1546}}       &$0.1445$       &\textit{\underline{0.1508}}       &$0.1289$       &\textit{\textbf{0.1220}} \\
\end{array}
\vspace{2em}


\begin{flushleft}
\noindent\textbf{Table S15. Performance metrics for predicted $\Delta B$ at Threshold = \textbf{450 nT}.}
\end{flushleft}

\begin{array}{c | c c c c|c c c c}
        Metric & SET A & SET B & SET C & SET D &
        SET E & SET F & SET G & SET H\\
        \hline
        POD     &\textit{\underline{0.4211}}       &$0.4243$       &$0.4589$       &\textit{\textbf{0.4803}}       &$0.5230$       &\textit{\underline{0.4901}}       &$0.5855$       &\textit{\textbf{0.6184}} \\
        POF     &$0.0487$       &\textit{\textbf{0.0482}}       &$0.0536$       &\textit{\underline{0.0601}}       &$0.0715$       &\textit{\textbf{0.0690}}       &$0.1127$       &\textit{\underline{0.1246}} \\
        FAR     &$0.2768$       &\textit{\textbf{0.2732}}       &$0.2791$       &\textit{\underline{0.2930}}       &\textit{\textbf{0.3117}}       &$0.3181$       &$0.3894$       &\textit{\underline{0.4003}} \\
        MR      &\textit{\underline{0.1552}}       &$0.1544$       &$0.1472$       &\textit{\textbf{0.1431}}       &$0.1343$       &\textit{\underline{0.1419}}       &$0.1236$       &\textit{\textbf{0.1163}} \\
\end{array}
\vspace{2em}

\end{article}
\clearpage